# Cryptography without Long-Term Quantum Memory and Global Entanglement

*Classical Setups for: One-Time Programs, Copy Protection, Stateful Obfuscation*

**Lev Stambler**

levstamb@umd.edu

Joint Center for Quantum Information and Computer Science, University of Maryland, College Park, MD 20742, USA

Department of Computer Science, University of Maryland, College Park, MD 20742, USA

NeverLocal Ltd. London, WC2H 9JQ, UK

April 21, 2025

## Abstract

We show how oracles which only allow for classical query access can be used to construct a variety of quantum cryptographic primitives which *do not* require long-term quantum memory or global entanglement. Specifically, if a quantum party can execute a semi-quantum token scheme (Shmueli 2022) with probability of success $\frac{1}{2} + \delta$, we can build powerful cryptographic primitives with a multiplicative logarithmic overhead for the desired correctness error. Our scheme makes no assumptions about the quantum party's noise model except for a simple independence requirement: noise on two sets of non-entangled hardware must be independent.

Using semi-quantum tokens and oracles which can only be queried classically, we first show how to construct a "short-lived" semi-quantum one-time program (OTP) which allows a classical sending party to prepare a one-time program on the receiving party's quantum computer. We then show how to use this semi-quantum OTP to construct a semi-quantum "stateful obfuscation" scheme (which we term "RAM obfuscation"). Importantly, the RAM obfuscation scheme does not require long-term quantum memory or global entanglement. Finally, we show how RAM obfuscation can be used to build long-lived one-time programs and copy-protection schemes.

## 1 Introduction

Quantum cryptography is a field that has seen rapid growth in the last few decades, enabling a panoply of new cryptographic primitives, impossible in the classical world. One of those primitives is the one-time program (OTP) [1], which allows a sender to prepare a quantum state on the receiver's side such that the receiver can evaluate a function *only once* using the quantum state. Some variants of OTPs, such as for sufficiently randomized functionalities, are possible to construct in the quantum setting [2–4], while others require stronger assumptions, mainly stateless oracles which only accept classical inputs [5–8] or computational restrictions on the receiver's quantum computational power [9–12]. We note that classically accessible oracles can be instantiated using a variety of different assumptions, such as TEEs, custom hardware, or even a trusted third party/ distributed system.

Unfortunately though, most OTP constructions and other quantum primitives require long-term quantum memory or global entanglement to capture their full power. For example, one would hope that a one-time program could be stored for a few days or weeks prior to its evaluation, but this is not possible in the current state of quantum technology. Specifically, in order to have a one-time program which can be stored for a long time, existing OTP schemes would require a sort of *quantum memory* which can last for a long time.

When assuming a simple noise model, such as Markovian noise, we can use quantum error correction with a logarithmic overhead in the number of qubits to store the OTP. However, given the current state of quantum technology, we know that simple noise models do not hold in practice, especially when considering longer time scales [13–16]. It is thus unclear whether we can achieve a logarithmic overhead in the number of qubits for long-term quantum memory in practice. Indeed, a detailed analysis by Gidney and Ekerå shows that quantum computers must have quite a large overhead in order to achieve reasonable logical error rates [17].

Moreover, when we consider other primitives, such as copy-protection [18], we run into even more constraints. Specifically, the existing schemes may require *global entanglement* to run the program. For example, existing instantiations of generalized copy-protection [19] require running the entire quantum state in superposition.

In this work, we assume the existence of classically accessible oracles and semi-quantum tokens [20] to construct a variety of quantum primitives which do not require long-term quantum memory or global entanglement. Moreover, we show that these primitives are "semi-quantum", meaning that the sender/ preparing party can be classical while the receiver/ evaluating party is quantum[1].

We first construct a semi-quantum OTP which allows a sender to prepare a quantum state on the receiver's side such that the receiver can evaluate a function using their quantum state. We then show how to use this semi-quantum OTP to construct a semi-quantum "stateful obfuscation" scheme (which we term "RAM obfuscation"). We can define RAM obfuscation as a scheme which prepares a black-box circuit on the receiver's side, $C_0$, such that when the receiver evaluates $C_0(x)$, the circuit $C_0$ transitions to a new circuit $C_1$ depending on the input $x$. Then, the receiver can evaluate $C_1$ on a new input $x'$ to get a new circuit $C_2$ and so on.

In a similar way to Goyal et. al's method of bootstraping one-time programs to reusable programs [21], we use our semi-quantum OTP to build RAM obfuscation. Our construction of RAM obfuscation can be thought of as a sort of "refreshing" of the semi-quantum OTP: i.e. we build RAM obfuscation by having a one-time program output a new one-time program upon evaluation. Importantly, because our one-time programs are semi-quantum, the preparation of the new one-time program does not require any entanglement with the old one-time program! Thus, we can view the new one-time program as only having a "classical" connection to the old one-time program. And so, our RAM obfuscation scheme does not require long-term quantum memory or global entanglement for the receiver to continuously evaluate the program. Using a simple Chernoff bound and our noise-independence assumption between non-entangled hardware, we can also show that the overhead of our scheme is logarithmic in the correctness error of the underlying semi-quantum token scheme.

---

[1] Technically, we also require non-malleable and CPA public key encryption schemes, though we skip over this detail in the introduction.

Finally, we show how our RAM obfuscation scheme can be used to build "long-lived" one-time programs and copy-protection schemes[2]. Long-lived one-time programs are simply a one-time program which can be stored for a long duration of time prior to its evaluation. Both constructions are relatively simple and follow from the construction of RAM obfuscation and its soundness properties. Because our RAM obfuscation scheme does not require long-term quantum memory or global entanglement, we immediately get that our long-lived one-time programs and copy-protection schemes do not require long-term quantum memory or global entanglement as well.

## 1.1 Main Results

We now give a high-level overview of our results.

**Fault-Tolerant Semi-Quantum Tokens**

First, we show how a semi-quantum token scheme [20] can be made "fault-tolerant" with minimal assumptions on the underlying quantum hardware. At a high level, a semi-quantum token scheme allows a classical party to prepare a quantum state on the receiver's side such that the receiver can "sign" a classical bit on behalf of the sending party using their quantum state. Importantly, the receiver should only be able to sign once.

To encode a semi-quantum token scheme, we require the following:
- The probability of a correctness error is less than $\frac{1}{2} - \delta$ for some $\delta \in \frac{1}{\text{poly}(\cdot)}$.
- We can run independent quantum circuits simultaneously and across time which do not impact each other in terms of noise/ error rate.

Then, we can simply construct a fault-tolerant semi-quantum token scheme by running the semi-quantum token scheme in parallel. A signature on a bit $b$ is accepted if a majority of the signatures are accepted for the underlying semi-quantum token schemes. Because all of the semi-quantum token schemes are independent, we can use a Chernoff bound to show that the probability of a correctness error decreases exponentially with the number of repeated semi-quantum token schemes.

Given that the remainder of our constructions (one-time programs, RAM obfuscation, etc.) are built on top of semi-quantum tokens, their fault-tolerance properties are inherited from the semi-quantum token scheme as well.

**Semi-Quantum One-Time Programs**

We will now give a high-level overview of our construction of semi-quantum one-time programs from semi-quantum tokens. We start by requiring that the semi-quantum token scheme has a single-round setup process. We note that the setup for the token scheme in Ref. [20] is already single-round. We use a classically accessible oracle as follows[3]:

---

[2]Given that RAM obfuscation implies black-box obfuscation, we can also use RAM obfuscation to build many other primitives. We also note that RAM obfuscation is on its own quite a powerful primitive.

[3]Within the scheme itself, the classical oracle is not instance specific but rather a "universal" oracle which can be used to evaluate any function. This universality is important for our construction of RAM obfuscation.

1. For each index $i \in [n]$, the receiver and sender engage in the setup process for the semi-quantum token scheme
2. The sender also sends a classical oracle which only allows for the evaluation of a function $f : \{0,1\}^n \to \{0,1\}^m$ on $x$ if each input bit of $x$ is *signed* under the sender's key
3. To evaluate the function $f$ on $x$, the receiver "signs" each bit of $x_i$ using their $i$-th semi-quantum token and then sends $x$ to the oracle as well as their signatures.

We can see that in the above, the receiver can only evaluate the function $f$ on $x$ once as the receiver can only sign for each $i \in [n]$ once. Moreover, the protocol has a single-round setup as the receiver sends the first message to the sender. We thus have our first result:

> *Theorem 1.1* (Informal Statement of Semi-Quantum One-Time Programs): Assuming the existence of semi-quantum token scheme and classically accessbile oracles, then we can construct a semi-quantum one-time program with a single-round setup.

**Semi-Quantum RAM Obfuscation**

Given a semi-quantum one-time program, we can then construct a semi-quantum RAM obfuscation scheme by having the one-time program output a new one-time program state upon evaluation. In more detail, say we want to obfuscate a recursive function $P_i(x)$ which outputs some output $y_i$ and its next evaluation function $P_{i+1}(\cdot)$. We can then encode a modified version, $\overline{P}_0$, into a one-time program such that when the receiver evaluates $\overline{P}_0(x)$, the program outputs $y_0$ as well as a new one-time program which encodes the next evaluation function $\overline{P}_1$. We can do this because our one-time programs have a single-round setup: the evaluating party can initiate the interactive setup process for $\overline{P}_1$ *prior to evaluating* $\overline{P}_0$ and so $\overline{P}_0$ can act as the "sending party" for $\overline{P}_1$[^4]!

We can follow this recursion to get a new one-time program $\overline{P}_i$ which encodes the next setup function for $\overline{P}_{i+1}$ as well. In this way, we can build RAM obfuscation by having a one-time program output a new one-time program upon evaluation. For a visualization of our RAM obfuscation scheme, see Figure 1.

Moreover, we have a rather straightforward property inherited from the semi-quantum one-time program: our RAM obfuscation scheme is fault-tolerant, single-round setup, and semi-quantum.

> *Theorem 1.2* (Informal Statement of RAM Obfuscation): Assuming the existence of a single-round setup semi-quantum token scheme, we can construct a semi-quantum RAM obfuscation scheme which also has a single-round setup.

We also get the following fault-tolerance property:

---

However, for the sake of simplicity, our introduction will focus on a specific instance of the oracle. Within the real protocol, the oracle itself is fixed as part of a global setup and is not instance specific.

[^4]: We note that within the actual protocol, the setup process for $\overline{P}_1$ requires some additional information prior to the receiver generating their setup messages. We can think of this additional information as being part of the setup process for $\overline{P}_0$.

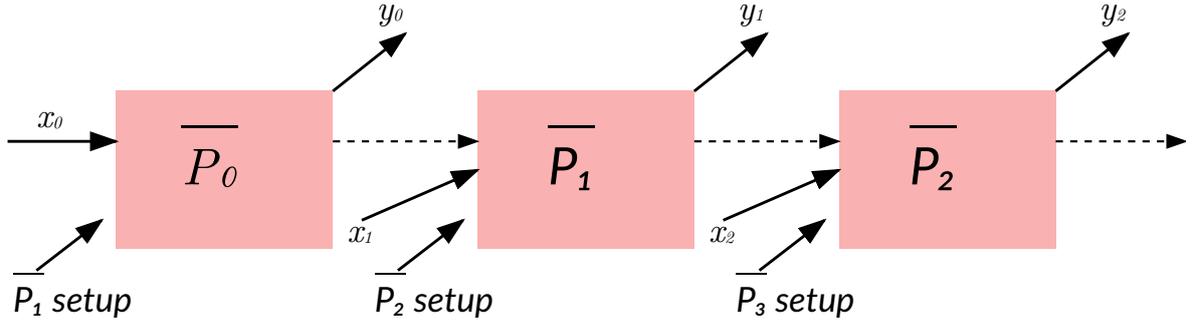

Figure 1: A visualization of our RAM obfuscation scheme. Each box, $\overline{P}_i$ represents a one-time program which outputs $y_i = P_i(x_i)$ and the next one-time program $\overline{P}_{i+1}$ given its setup messages.

> *Theorem 1.3* (Informal Statement of Fault-Tolerance for RAM Obfuscation): Assuming the existence of a semi-quantum token scheme with correctness error less than $\frac{1}{2} - \delta$ and classically accessbile oracles, we can construct a semi-quantum RAM obfuscation scheme with input size $n$ which has correctness error $\varepsilon$ for $\ell$ evaluations of the program with multiplicative overhead $O\big(\log\big(\frac{\ell n}{\varepsilon}\big) \cdot \frac{1}{\delta^2}\big)$.

To highlight the difference between our fault-tolerance theorem and existing fault-tolerance theorems, we provide a simple example of a potential architecture which could be used to build our fault-tolerant semi-quantum token scheme (and thus our RAM obfuscation scheme) in Figure 2.

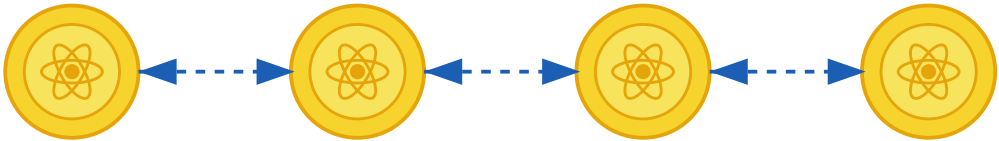

Figure 2: Each quantum computer is represented by a yellow circle and *classical* communication is represented by a dashed blue line.

**Long-Lived One-Time Programs and Copy-Protection**

Finally, we get to our two main applications of RAM obfuscation explored in this paper: long-lived one-time programs and copy-protection schemes. We note that both of these applications are built on top of our RAM obfuscation scheme and thus inherit its fault-tolerance properties and lack of long-term quantum memory and global entanglement.

First, we can use RAM obfuscation to build "long-lived" one-time programs. The construction is a simple application of RAM obfuscation. To encode a function $f$, we can simply encode the function into a recursive program $P_i$ such that $P_i(\bot)$ outputs nothing except for $P_{i+1}(\cdot)$ and $P_i(x)$ outputs $f(x)$ and next recursive program $P_{i+1}^{\bot}$ where $P_{i+1}^{\bot}$ is a recursive encoding of a "null" function. We can then see that the receiver can choose to evaluate $f$ on $x$ at any-time step but can only do so once. Because each of the one-time program evaluations are semi-quantum, no entanglement is needed between the evaluations. We thus have the following result:

> *Theorem 1.4* (Informal Statement of Long-Lived One-Time Programs): Assuming the existence of a semi-quantum token scheme with correctness error less than $\frac{1}{2}$ and classically accessbile oracles, we can construct a long-lived and fault-tolerant one-time program with a logarithmic overhead in the correctness error.

We then show how to use RAM obfuscation to build semi-quantum copy-protection schemes. To copy-protect function $f$, we first choose some secret PRK key $K$ and then encode the function into a recursive program $P_i(x, k_i)$ such that $P_i(x, k_i)$ outputs $(f(x), \text{PRF}(K, i+1))$ and the next recursive program $P_{i+1}(\cdot, k_{i+1})$ if and only if $k_i = \text{PRF}(K, i)$. Otherwise, the program outputs nothing except for $P^\perp_{i+1}(\cdot, k_{i+1})$ where $P^\perp_{i+1}$ is a recursive encoding of a "null" function, effectively destroying the program.

Then, by the soundness of RAM obfuscation, we cannot evaluate $P_i$ twice for the same $i$. But then, any non-communicating parties which try to clone $f$ have to do one of the following:
1. Evaluate $P_i$ twice for the same $i$. Note that this breaks the soundness of RAM obfuscation and is thus impossible.
2. Evaluate $P_i$ once and then the other party evaluates $P_j$ for some $j \neq i$. Assuming that $i < j$, the second party must have known $k_i$ in order to evaluate $P_{i+1}, ...P_{j-1}$. Also, the second party must have been able to evaluate $P_{j-1}$ in order to get $k_j$. But then, the second party must have evaluated $P_i$ at some point in order to get $k_i$, breaking the soundness of RAM obfuscation!

Also, just as with the long-lived one-time programs, no entanglement is needed between the evaluations.

We thus have the following result:

> *Theorem 1.5* (Informal Statement of Copy-Protection): Assuming the existence of a semi-quantum token scheme (with correctness error $< \frac{1}{2}$) and classically accessible oracles, we can construct a semi-quantum copy-protection scheme which is also also fault-tolerant with a logarithmic overhead in the correctness error.

## Notation

Throughout this paper, we will use several notational conventions. Generally, we will use lowercase letters for vectors and uppercase letters for matrices. When referring to the elements of a vector or matrix, we will use superscript notation, e.g. $x^i$ refers to the $i$th element of vector $x$ and $A^i$ to refer to the $i$-th row of matrix $A$. For sets, we will generally use script letters such as $\mathcal{S}$, and for any set $\mathcal{S}$, $|\mathcal{S}|$ will denote its cardinality. We will also use calligraphic letters to denote oracles, usually denoted with $\mathcal{O}$. We will also use the notation $\mathcal{A}^\mathcal{O}$ to denote an algorithm $\mathcal{A}$ with *classical* oracle access to $\mathcal{O}$. Also, for any function $f$, $f(\cdot)$ represents its evaluation.

We will also use the notation 1-$P$ to denote an oracle which can only be queried classically once. And so, $\mathcal{A}^{1\text{-}P}$ denotes an algorithm $\mathcal{A}$ with oracle access to $P$ which can only be classically queried once.

**Outline**

The paper is organized as follows. In Section 2, we provide a brief overview of the necessary background material for the rest of the paper. Then in Section 3, we review semi-quantum tokenized signature schemes and show how they can be made fault-tolerant in a simple way. Next, in Section 4, we define and construct receiver-first single-round semi-quantum one-time programs (OTP). Following our OTP construction, we construct a semi-quantum RAM obfuscation scheme in Section 5. Then, in Section 6, we show how to use our RAM obfuscation scheme to construct long-lived OTPs and copy-protection schemes. Finally, in Section 7, we conclude with a discussion of future work and open problems.

## 2 Preliminaries

In this section, we provide a brief overview of the necessary background material for the rest of the paper.

### 2.1 Different Notions of Public Key Encryption

We will use the simulation-based definition of non-malleable public key encryption from Bellare et al. [22].

*Definition 2.1 (CPA PK Non-Malleable Encryption)*: Let PK = (PK.Gen, PKEncr, PK.Decr) be a public key encryption scheme. Let $\mathcal{A} = (\mathcal{A}_1, \mathcal{A}_2)$ be an adversary consisting of a pair of BQP algorithms. Let Sim = $(\text{Sim}_1, \text{Sim}_2)$ be a pair of algorithms which we call the simulator. We say that a PK scheme is non-malleable for every (quantum) adversary, $\mathcal{A}$, and polynomial time computable function, Sim, if for all relations $R$,

$$\Pr\big[\text{Expt}_{\mathcal{A},\,\text{PK},R}(\lambda) = 1\big] - \Pr\big[\text{Expt}_{\text{Sim},\,\text{PK},R}(\lambda) = 1\big] \leq \text{negl}(\lambda) \qquad (1)$$

where the experiments are defined as

$$
\begin{array}{ll}
\text{Expt}_{\mathcal{A},\,\text{PK},R}(\lambda): & \text{Expt}_{\text{Sim},\,\text{PK},R}(k) \\
\quad (\text{pk},\text{sk}) \leftarrow \text{PK.Gen}(1^\lambda) & \quad (\text{pk},\text{sk}) \leftarrow \text{PK.Gen}(1^k) \\
\quad (M,s_1,s_2) \leftarrow \mathcal{A}_1(\text{pk}) & \quad (M,s_1,s_2) \leftarrow \text{NMSim}_1(\text{pk}) \\
\quad x \leftarrow M;\, y \xleftarrow{\$} \text{PK.Encr}_{\text{pk}}(x) & \quad x \leftarrow M \\
\quad y' \leftarrow \mathcal{A}_2(y,s_2) & \quad y' \leftarrow \text{NMSim}_2(s_2) \\
\quad x' \leftarrow \text{PK.Decr}_{\text{sk}}(y') & \quad x' \leftarrow \text{PK.Decr}_{\text{sk}}(y') \\
\quad \text{If } R(x,x',M,s_1) \text{ then return 1} & \quad \text{If } R(x,x',M,s_1) \text{ then return 1} \\
\quad \text{Else return 0} & \quad \text{Else return 0} \qquad (2)
\end{array}
$$

We also use the CPA notion of secure public key encryption from Ref. [23]:

*Definition 2.2 (CPA PK Encryption)*: Let PK = (PK.Gen, PKEncr, PK.Decr) be a public key encryption scheme. We say that a PK scheme is CPA secure for every (quantum) polynomial time computable function, $\mathcal{A}$, and polynomial time computable function, Sim, if

$$\Pr\big[\text{Expt}_{\mathcal{A},\,\text{PK},0}(\lambda) = 1\big] - \Pr\big[\text{Expt}_{\mathcal{A},\,\text{PK},1}(\lambda) = 1\big] \leq \text{negl}(\lambda) \qquad (3)$$

where the experiments are defined as

$$
\begin{array}{ll}
\texttt{Expt}_{\mathcal{A},\ \mathsf{PK},0}(\lambda): & \texttt{Expt}_{\mathcal{A},\ \mathsf{PK},1}(\lambda): \\
\quad (\mathsf{pk},\mathsf{sk}) \leftarrow \mathsf{PK.Gen}(1^\lambda) & \quad (\mathsf{pk},\mathsf{sk}) \leftarrow \mathsf{PK.Gen}(1^\lambda) \\
\quad (m_{1,0}, m_{1,1}, ..., m_{q,0}, m_{q,1}) \leftarrow \mathcal{A}_1(\mathsf{pk}) & \quad (m_{1,0}, m_{1,1}, ..., m_{q,0}, m_{q,1}) \leftarrow \mathcal{A}_1(\mathsf{pk}) \\
\quad \text{where } |m_{i,0}| = |m_{i,1}| \text{ for } i \in [q] & \quad \text{where } |m_{i,0}| = |m_{i,1}| \text{ for } i \in [q] \\
\quad \text{For all } i \in [q],\ c_i \leftarrow \mathsf{PK.Encr}_{\mathsf{pk}}(m_{i,0}) & \quad \text{For all } i \in [q],\ c_i \leftarrow \mathsf{PK.Encr}_{\mathsf{pk}}(m_{i,1}) \\
\quad \hat{b} \leftarrow \mathcal{A}(c_1, ..., c_q) \text{ for } \hat{b} \in \{0,1\} & \quad \hat{b} \leftarrow \mathcal{A}(c_1, ..., c_q) \text{ for } \hat{b} \in \{0,1\} \\
\quad \text{Return } \hat{b} & \quad \text{Return } \hat{b}
\end{array}
\qquad (4)
$$

## 2.2 Entropy and Pseudo-Entropy

We also make use of basic notions of information theory and its computational analogues.

*Definition 2.3 (Minimum Entropy)*: The minimum entropy of a random variable $X$, conditioned on $Y$, is defined as

$$H_\infty(X \mid Y) = -\log_2 \mathbb{E}_{y \leftarrow Y}\left(\max_{\{x\}} \Pr[X = x \mid Y = y]\right) \qquad (5)$$

*Definition 2.4 (HILL Entropy [24,25])*: Let $X, Y$ be ensembles of jointly distributed random variables. We define the pseudo-entropy of $X$ conditioned on $Y$ to be at least $\ell(\lambda)$, denoted by $H_{\mathsf{HILL}}(X \mid Y) \geq \ell(\lambda)$ if there exists some $X'$ distributed with $Y$ such that $(X,Y) \stackrel{c}{\approx} (X',Y)$ and $H_\infty(X' \mid Y) \geq \ell(\lambda)$.

# 3 Semi-Quantum Tokenized Signature Schemes

We present Shmueli's semi-quantum tokenized signature scheme [20] though we slightly modify the definition in the following ways; we:

- split up the classical communication protocol into two separate algorithms. First, we have a fixed secret/ public key pair for the classical party and a separate *evaluation key* for a single quantum token.
- omit the quantum verification algorithm $\mathsf{QV}$ from the definition, as it is not used in our construction.
- relax correctness to hold with probability $\frac{1}{2} + \delta$ instead of $1 - \mathsf{negl}(\lambda)$ for $\delta \in \mathsf{poly}^{-1}(\lambda)$.

We note that our changes are quite surface-level and do not change the underlying construction or security of the scheme proposed by Shmueli.

*Definition 3.1 (Semi-quantum tokens, CQ-TOK [20])*: A semi-quantum tokenized signature scheme consists of algorithms $(\mathsf{Sen}, \mathsf{Rec}, \mathsf{Sign}, \mathsf{CV})$ with the following syntax:

- $\mathsf{pk} \leftarrow \mathsf{Setup}(\mathsf{sk})$: A classical polynomial-time deterministic algorithm that takes as input a classical secret key $\mathsf{sk}$ and outputs a public key $\mathsf{pk}$.
- $(\mathsf{ek}, |\mathsf{qt}\rangle_{\mathsf{ek}}) \leftarrow \langle \mathsf{Sen}(\mathsf{sk}, \mathsf{pk}), \mathsf{Rec}(\mathsf{pk}) \rangle_{\mathrm{OUT}_{\mathsf{Sen}}, \mathrm{OUT}_{\mathsf{Rec}}}$: a classical-communication protocol between a PPT algorithm Sen and a QPT algorithm Rec. At the end of interaction the sender outputs a classical evaluation key $\mathsf{ek}$ and the receiver outputs a quantum state $|\mathsf{qt}\rangle_{\mathsf{ek}}$.
- $\sigma_b \leftarrow \mathsf{Sign}(\mathsf{pk}, \mathsf{ek}, |\mathsf{qt}\rangle_{\mathsf{ek}}, b)$: A QPT algorithm that gets as input the public key $\mathsf{pk}$, a candidate token $|\mathsf{qt}\rangle_{\mathsf{pk}}$ and a bit $b \in \{0,1\}$ and outputs a classical string $\sigma_b$.
- $\mathsf{CV}(\mathsf{pk}, \mathsf{ek}, \sigma_b, b) \in \{0,1\}$: A classical polynomial-time deterministic algorithm that takes as input the public key $\mathsf{pk}$, evaluation key $\mathsf{ek}$, a classical string $\sigma_b$ and a bit $b \in \{0,1\}$, and outputs a bit.

The scheme satisfies the following guarantees:

**Statistical Correctness:** There exists a $\delta \in \text{poly}^{-1}(\lambda)$ such that for every $\lambda \in \mathbb{N}$,
$$\Pr[\texttt{CV}(\texttt{pk}, \texttt{ek}, \texttt{Sign}(\texttt{ek}, |\texttt{qt}\rangle_{\texttt{ek}}, b), b)] \geq \frac{1}{2} + \delta \qquad (6)$$
where the probability is over $\left(\texttt{ek}, |\texttt{qt}\rangle_{\texttt{pk}}\right) \leftarrow \langle \texttt{Sen}, \texttt{Rec}\rangle_{\text{OUT}_{\text{Sen}}, \text{OUT}_{\text{Rec}}}$)

**Security:** For every $\mathcal{A} = \{A_\lambda, \rho_\lambda\}_{\lambda \in \mathbb{N}}$ a quantum polynomial-time algorithm there exists a negligible function $\text{negl}(\cdot)$, such that for states sampled by the setup, $\texttt{pk} \leftarrow \texttt{Setup}$ and subsequent interaction with the sender, $(\texttt{ek}, \text{QT}) \leftarrow \langle \texttt{Sen}, \mathcal{A}\rangle_{\text{OUT}_{\text{Sen}}, \text{OUT}_A}$, for every $\lambda \in \mathbb{N}$, then the probability that
$$\text{QT} = (\sigma_0, \sigma_1), \text{ such that } \texttt{CV}(\texttt{pk}, \sigma_0, 0) = \texttt{CV}(\texttt{pk}, \sigma_1, 1) = 1 \qquad (7)$$
is negligible in $\lambda$.

We further require that the protocol $\langle \texttt{Sen}, \texttt{Rec}\rangle_{\text{OUT}_{\text{Sen}}, \text{OUT}_{\text{Rec}}}$ is **single-round** protocol. More formally, we have the following definition:

*Definition 3.2 (Receiver-first single-round CQ-TOK):* A CQ-TOK scheme is receiver-first single-round if the protocol $\langle \texttt{Sen}, \texttt{Rec}\rangle_{\text{OUT}_{\text{Sen}}, \text{OUT}_{\text{Rec}}}$ is a single-round protocol. Specifically, the protocol is initiated by the receiver sending a message to the sender, and the sender responds with a message to the receiver. When a CQ-TOK is receiver-first single-round, we specify $\texttt{Sen}(\texttt{sk}, \texttt{pk}, z)$ as the sender's algorithm where $z$ is the receiver's message instead of as an interactive protocol. Moreover, as Sen is the last algorithm, we also require that the message of Sen is simply just the evaluation key $\texttt{ek}$. This is illustrated in Table 1.

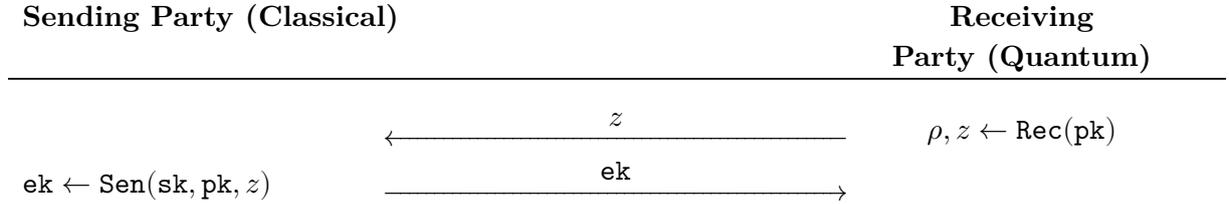

Table 1: Illustration of receiver-first single-round for the CQ-TOK scheme

For completeness, we present the modified setup and interactive portion of the scheme in Protocol 8 within Appendix A though we note that the scheme is essentially the same as the original scheme proposed by Shmueli. As classical verification and signing remain unchanged, we do not present them in Appendix A.

## 3.1 Fault-Tolerant Lifting for CQ-TOK Schemes

We now highlight one of the main advantages for the use of CQ-TOK schemes in our construction. Mainly, we show that the CQ-TOK scheme can be lifted to a fault-tolerant version of the scheme as long as the *correctness* of the scheme holds with probability $1/2 + \delta$ for $\delta = \text{poly}^{-1}(\lambda)$.

> *Theorem 3.1* (Fault-tolerant lifting of CQ-TOK): Let $(\texttt{Setup}, \texttt{Sen}, \texttt{Rec}, \texttt{CV})$ be a CQ-TOK scheme such that the following holds:
> - The scheme is correct with probability $1/2 + \delta$.
> - The scheme is secure against quantum adversaries.

> Then there exists a fault-tolerant version of the scheme $(\mathtt{Setup}', \mathtt{Sen}', \mathtt{Rec}', \mathtt{CV}')$ such that the correctness of the scheme holds with probability $1 - \varepsilon_{\mathrm{TOK}}$ for any $\varepsilon_{\mathrm{TOK}} > 0$ with multiplicative overhead of $O\!\left(\log\!\left(\frac{1}{\varepsilon_{\mathrm{TOK}}}\right) \cdot \frac{1}{\delta^2}\right)$ over the original scheme.

*Proof sketch*: The construction follows from a simple repetition of the CQ-TOK scheme. Using $w = O\!\left(\log\!\left(\frac{1}{\varepsilon}\right) \cdot \frac{1}{\delta^2}\right)$ independent copies of the scheme, we can amplify the correctness of the scheme to $1 - \varepsilon$ using standard Chernoff bounds by requiring a signature for 0 to have at least $\frac{w}{2} + 1$ copies from the underlying scheme accept a signature on 0. To sign for 1, at least $\frac{w}{2} + 1$ copies of the scheme accept a signature for 1. The security of the scheme follows from the fact that the CQ-TOK scheme is secure against quantum adversaries and so producing a signature for 0 and 1 would require the adversary to produce a signature for both 0 and 1 for at least one copy of the scheme by the pigeonhole principle. ∎

# 4 Semi-Quantum One-Time Programs

In this section, we define a semi-quantum one-time program (C-OTP) with global setup and construct a semi-quantum one-time program from semi-quantum tokens (CQ-TOK) and classically accessible oracles.

*Definition 4.1 (Semi-quantum one-time program)*: A semi-quantum one-time program (C-OTP) with global setup outputting oracle $\mathcal{O}_{\text{C-OTP}}$ is defined by the tuple $(\mathtt{C\text{-}OTP.global\_setup}, \mathtt{C\text{-}OTP.setup}^{\mathcal{O}_{\text{C-OTP}}}, \mathtt{C\text{-}OTP.gen}^{\mathcal{O}_{\text{C-OTP}}}, \mathtt{C\text{-}OTP.eval}^{\mathcal{O}_{\text{C-OTP}}})$ such that for global security parameter $\lambda$ and setup, we have

$\mathtt{C\text{-}OTP.global\_setup(msk)}$: outputs global auxiliary information $\mathtt{aux}$ for master secret key $\mathtt{msk}$ as well as oracle $\mathcal{O}_{\text{C-OTP}}$. Then, for a fixed program $P$, we have
1. $\mathtt{C\text{-}OTP.setup}$ outputs a classical public key $\mathtt{pk}$ associated to $\mathtt{sk}$.
2. $\mathtt{C\text{-}OTP.gen}$ is defined by an interactive protocol, $\langle \mathcal{C}_{\text{C-OTP}}, \mathcal{Q}_{\text{C-OTP}} \rangle_{\mathrm{OUT}_{\mathcal{C}_{\text{C-OTP}}}, \mathrm{OUT}_{\mathcal{Q}_{\text{C-OTP}}}}$, where $\mathcal{C}_{\text{C-OTP}}$ is the classical party and $\mathcal{Q}_{\text{C-OTP}}$ is the quantum party. Moreover, we require that the protocol is single-round as pictured in Table 2. We have $\mathrm{OUT}_{\mathcal{Q}_{\text{C-OTP}}} = (|\!\!\nearrow\rangle, \mathtt{tag}, \mathtt{ek})$.
3. $\mathtt{C\text{-}OTP.eval}(x, |\!\!\nearrow\rangle, \mathtt{tag}, \mathtt{ek}, \mathtt{pk})$ outputs the evaluation of the program $P(x)$ on input $x$ given the state $|\!\!\nearrow\rangle$, tag $\mathtt{tag}$, and the evaluation key $\mathtt{ek}$.

| Program Delegating Party (Classical) | | Program Receiving Party (Quantum) |
|---|---|---|
| | $\xleftarrow{\quad \mathtt{tag} \quad}$ | $|\!\!\nearrow\rangle, \mathtt{tag} \leftarrow \mathcal{Q}_{\text{C-OTP}}(\mathtt{pk})$ |
| $\mathtt{ek}, \mathtt{pk}' \leftarrow \mathcal{C}_{\text{C-OTP}}(P, \mathtt{sk}, \mathtt{tag})$ | $\xrightarrow{\quad \mathtt{ek} \quad}$ | |

Table 2: Shape of C-OTP.gen

For soundness, we will use a simulation-based definition and leave the UC-based definition for future work. Because the interactive protocol within generation is single-round, we can define the soundness of the scheme relative to a classical oracle, $\mathcal{C}_{\text{C-OTP}}$, which can only be called once. In other words, we represent the sender's generation algorithm as a classical oracle which can only be queried once. This simplifies the soundness definition and our proofs.

*Definition 4.2 (`C-OTP` Simulation-Based Soundness)*: We say that a `C-OTP` scheme is sound with correctness error $\varepsilon_{\text{corr}}$ for program $P$ relative to `sk` and auxiliary information, `aux`, if:
- $H_{\mathsf{HILL}}(\mathtt{sk} \mid P, \mathtt{aux}) = H_{\mathsf{HILL}}(\mathtt{sk})$: i.e. the program $P$ is computationally independent of the secret key `sk`.
- For every adversary, $\mathcal{A}$, there exists a simulator Sim such the following holds:

$$\mathcal{A}^{\mathcal{O}_{\text{C-OTP}},\ 1\text{-}\mathcal{C}_{\text{C-OTP}}(P, \mathtt{sk}, \cdot)}(\mathtt{pk}, \mathtt{aux}) \stackrel{c}{\approx} \mathrm{Sim}^{\mathcal{O}_{\text{C-OTP}},\ 1\text{-}P}(\mathtt{aux}) \tag{8}$$

where 1-$P$ returns an evaluation of $P$ with probability $1 - \varepsilon_{\text{corr}}$.

## Instantiating Semi-Quantum One-Time Programs from `CQ-TOK`s

In this section, we will instantiate semi-quantum one-time programs (Definition 4.1) from semi-quantum tokens (`CQ-TOK`, as per Definition 3.1) as well as classically accessible oracles.

### 4.1 `C-OTP`s from a Classically Accessible Oracles and `CQ-TOK`s

Though one often constructs one-time memories to build one-time programs, we will construct one-time programs directly for ease of exposition[5].

The construction roughly follows the ideas presented in the introduction, though with one key difference:
1. We use a *global classically accessible oracle* which can be reused for multiple one-time programs.
2. The oracle has an associated master secret and public key (`msk, mpk`) which is used by the sending party to encrypt the one-time program alongside the evaluation key for the quantum tokens.
3. The classically accessible oracle only evaluates the program if the input has an associated signature.

Moreover, in the first step of the soundness proof, we make use of a *non-malleable* public-key encryption scheme which "ties" the encryption of the evaluation key to the encryption of the program: we can then ensure that any query to the oracle, attempting to evaluate the program, under a different evaluation key will fail. Then, we use make use of the "sign-once" nature of the quantum tokens to ensure that the evaluation key is only valid for a single evaluation of the program. Next, we use standard notions of public-key encryption to replace the encryption of the program by a randomly chosen program.

We now present the construction, with the classical oracle specified in Oracle 1 and the protocol specified in Protocol 2.

---

[5]To the author's knowledge, it may still be possible to first construct one-time memories and then construct one-time programs from one-time memories. But, because we use non-standard security definitions, this paper would still need to prove soundness of one-time programs from one-time memories. We thus find it easier to construct one-time programs directly rather than construct one-time memories and then construct one-time programs from one-time memories.

**Oracle 1:** Classically Accessible Oracle for the One-Time Program with secret msk

**Inputs** $(x, \text{ct}_A, \sigma)$ for input $x$, cipher-text $\text{ct}_A$ for $(P, \text{ek}, \text{pk})$, and signatures $\sigma$ for each bit of $x$.
**Hard-coded** msk is the master secret key

1 Use msk to decrypt $\text{ct}_A$ and obtain $\text{ek}, \text{pk}$ and $P$
2 Parse $\text{ek} = (\text{ek}_1, ..., \text{ek}_n)$
3 Parse $\text{pk} = (\text{pk}_1, ..., \text{pk}_n)$
4 **for** $i \in [n]$
5    Let $b_i$ be the $i$-th bit of $x$
6    **If** CQ-TOK.CV$(\text{pk}_i, \text{CQ-TOK.ek}_i, \sigma_i, b_i) = 0$
7       **Return** $\bot$
8 **Return** $P(x)$

---

**Protocol 2:** C-OTP from CQ-TOK for program $P : \{0,1\}^n \to \{0,1\}^n$

C-OTP.global_setup(msk) :

  Let $\mathcal{O}_{\text{C-OTP}}$ where $\mathcal{O}_{\text{C-OTP}}$ is the classical oracle for Oracle 1 with hard-coded master-secret key msk
  Let $\text{mpk} \leftarrow \text{PK.Gen}(\text{msk})$ be a public key for msk
  **Publish** auxiliary information $\text{aux}_{\text{C-OTP}} = (\text{mpk}, \mathcal{O}_{\text{C-OTP}})$

C-OTP.setup(sk) where $\text{sk} = (\text{sk}_1, ..., \text{sk}_n)$:

  $\text{pk}_i \leftarrow \text{CQ-TOK.Setup}(\text{sk}_i)$ for $i \in [n]$
  **Send**
$$\text{pk} = (\text{pk}_1, ..., \text{pk}_n) \qquad (9)$$
  to the receiver

C-OTP.gen :

  Follows the protocol outlined in Table 2 with the quantum (receiving) party $\mathcal{Q}_{\text{C-OTP}}$, first doing the following:
    **For** $i \in [n]$, let $|🗝\rangle_i, z_i \leftarrow \text{CQ-TOK.Rec}(\text{pk}_i)$ for classical tag $z_i$
    **Sends** $z = (z_1, ..., z_n)$ to the classical party
  The classical (sending) party upon receiving $z = (z_1, ..., z_n)$: $\mathcal{C}_{\text{C-OTP}}(P, \text{sk}, z)$ proceeds:
    Sample evaluation key $\text{CQ-TOK.ek}_i \leftarrow \text{CQ-TOK.Sen}(\text{pk}_i, \text{sk}_i, z_i)$
    Let $\text{CQ-TOK.ek} = (\text{ek}_1, ..., \text{ek}_n)$
    **Send** $\text{ct}_A = \text{PK.Encr}(\text{mpk}, [P, \text{pk}, \text{ek}])$ to the quantum party

C-OTP.eval$^{\mathcal{O}_{\text{C-OTP}}}(\text{pk}, x, \text{ct}_A, \text{CQ-TOK.ek}, |🗝\rangle)$

  Parse $\text{pk} = (\text{pk}_1, ..., \text{pk}_n)$ and $|🗝\rangle = |🗝\rangle_1, ..., |🗝\rangle_n$
  **For** $i \in [n]$, let $\sigma_i \leftarrow \text{CQ-TOK.Sign}(\text{pk}_i, \text{CQ-TOK.ek}_i, |🗝\rangle_i, x_i)$ and $\sigma = (\sigma_1, ..., \sigma_n)$
  **Output** $\mathcal{O}_{\text{C-OTP}}(x, \sigma, \text{ct})$

> *Theorem 4.1* (Soundness of `C-OTP` from `CQ-TOK`): Assuming a `CQ-TOK` scheme has correctness error $\varepsilon_{\text{TOK}}$, non-malleability and CPA security for public key encryption, and a classically accessible oracle, $\mathcal{O}_{\text{C-OTP}}$, the protocol Protocol 2 is sound with correctness error $n \cdot \varepsilon_{\text{TOK}}$ by the definition of Definition 4.2 as long as $H_{\text{HILL}}(\text{sk} \mid P, \text{aux}) = H_{\text{HILL}}(\text{sk})$.

*Proof*: We prove Theorem 4.1 by providing a simulator Sim such that the experiments are indistinguishable for any given adversary. We show that the real and simulated experiments are indistinguishable via the following hybrids.

- $\text{Hyb}_0$: the real protocol
- $\text{Hyb}_{1,0}$: replace the first call to $\mathcal{O}_{\text{C-OTP}}$ with $\mathcal{O}_1(x, \text{ct}', \sigma)$ as follows:
  ▸ If $P'$ is independent of $P$ (i.e. $H_{\text{HILL}}(P \mid P') = H_{\text{HILL}}(P)$ and vice-versa), return $\mathcal{O}_{\text{C-OTP}}(x, \text{ct}', \sigma)$
  ▸ If $\text{ct}' = \text{ct}_A$, return $\mathcal{O}_{\text{C-OTP}}(x, \text{ct}_A, \sigma)$
  ▸ Otherwise, return $\perp$
- $\text{Hyb}_{1,i}$: replace the $i$-th call to $\mathcal{O}_{\text{C-OTP}}$ with $\mathcal{O}_1\big(x, \text{ct}'_{P',\text{pk}',\text{ek}'}, \sigma\big)$ as follows:
  ▸ If $P'$ is independent of $P$ given the prior $i-1$ calls to $\mathcal{O}_1$, return $\mathcal{O}_1(x, \text{ct}', \sigma)$
  ▸ If $\text{ct}' = \text{ct}_A$, return $\mathcal{O}_{\text{C-OTP}}(x, \sigma, \text{ct}_A, \text{ek}')$
  ▸ Otherwise, return $\perp$
- $\text{Hyb}_2$: Replace $\text{ct}_A = \text{ct}_{P, \text{sk}}$ with $\text{ct}_\perp = \text{PK.Encr}(\text{mpk}, [\perp, \text{pk}, \text{ek}])$. Then, replace $\mathcal{O}_1(x, \text{ct}', \sigma)$ with $\mathcal{O}_2(x, \text{ct}', \sigma)$ as follows:
  ▸ If $\text{ct}' = \text{ct}_\perp$, return $\mathcal{O}_1(x, \text{ct}_A, \sigma)$
  ▸ Otherwise, return $\mathcal{O}_1(x, \text{ct}', \sigma)$
- $\text{Hyb}_3$: Replace $\mathcal{O}_2$ with $\mathcal{O}_3^{1\text{-}P}(x, \text{ct}', \sigma)$ as follows:
  ▸ If $\text{ct}' \neq \text{ct}_A$, return $\mathcal{O}_2(x, \text{ct}', \sigma)$
  ▸ Check if $\text{CQ-TOK.CV}(\text{pk}_{\text{CQ-TOK}}^i, \text{ek}_i, \sigma_i, x_i) = 1$ for all $i \in [n]$.
    – If the check passed, then check if $x$ has been called and if yes, retrieve $P(x)$ from memory. If not, return $P(x)$ and store $P(x)$. Note that each check passes with probability at least $1 - \varepsilon_{\text{corr}}$ and thus, for honest generation, the check passes with probability $1 - n \cdot \varepsilon_{\text{corr}}$.
    – Otherwise, return $\perp$
- $\text{Hyb}_4$: The same as before except that the simulator samples random public key and private key pairs for the quantum tokens, $\widehat{\text{pk}}_i, \widehat{\text{sk}}_i$. Then sample a corresponding evaluation key, $\widehat{\text{ek}}_i$, for the tag. Replace the encryption of $\text{ct}_\perp$ with $\widehat{\text{ct}}_\perp = \text{PK.Encr}\big(\text{mpk}, \big[\perp, \widehat{\text{pk}}, \widehat{\text{ek}}\big]\big)$
- $\text{Hyb}_5$: The simulated protocol.

■

For a proof of the indistinguishability of the hybrids, we refer the reader to Appendix B.

Finally, as a simple corollary of the correctness error and overhead of the `CQ-TOK` scheme, we have the following:

**Corollary 4.1** *(Fault-Tolerance Overhead)*: To achieve a correctness error of $1 - \varepsilon_{\text{corr}}$ for the `C-OTP` scheme, we require $O\big(\log(n/\varepsilon_{\text{corr}}) \cdot \frac{1}{\delta^2}\big)$ copies of the `CQ-TOK` scheme and thus incur a multiplicative overhead of $O\big(\log(n/\varepsilon_{\text{corr}}) \cdot \frac{1}{\delta^2}\big)$ over just using a single `CQ-TOK` scheme per input bit.

# 5 RAM-Oracles

We can view the "refreshing" of one-time programs as a sort of *one-time to many-time* lift as in Ref. [7]. Unlike the normal notion of stateful oracles though, we will consider a slightly stronger notion. To disambiguate, we will refer to our notion of stateful oracles as RAM-oracles rather than stateful oracles. The motivation for this is that we provide a mechanism for a (stateless) oracle to be equipped with some "encrypted and authenticated" RAM of fixed size.

*Definition 5.1 (RAM-Oracle)*: A stateful RAM-oracle for a program $P$, connoted $R\mathcal{O}$, with authenticated RAM, is an oracle which maintains an updatable state across queries. Specifically, we model any algorithm $\mathcal{A}$ with oracle access to $R\mathcal{O}_p$ and starting state $\text{RAM}_0$ as follows: the algorithm's queries to the oracles, $x_1, ..., x_\ell$ with outputs $y_1, ..., y_\ell$, can be broken down as follows
$$(y_i, \text{RAM}_i) = P(x_i, \text{RAM}_{i-1}).$$
where $\text{RAM}_i$ is the updated state of the oracle after query $x_i$.

We then say that a scheme is a RAM-blackbox obfuscator if it can be used to construct a RAM-oracle for any program $P$. Specifically, we use the following definition:

*Definition 5.2 (RAM-Blackbox Obfuscator)*: A RAM-blackbox obfuscator is a tuple of algorithms $(R\mathcal{O}.\texttt{send}, R\mathcal{O}.\texttt{eval})$ such that:
- $R\mathcal{O}.\texttt{send}(1^\lambda, P, \text{RAM}_0)$ specifies an interactive protocol between the sending, $R\mathcal{O}.\mathcal{S}$ and receiving party, $R\mathcal{O}.\mathcal{R}$.
- $R\mathcal{O}.\texttt{eval}\left(\tilde{P}, x, \widetilde{\text{RAM}}_{i-1}\right)$ outputs the evaluation of the program $P(x)$ on input $x$ and new memory state $\widetilde{\text{RAM}}_i$ given the state $\widetilde{\text{RAM}}_{i-1}$.

Then, for soundness, we require that for every BQP adversary, $\mathcal{A}$, there exists a simulator $\text{Sim}_{R\mathcal{O}}$ such that
$$\mathcal{A}\big(R\mathcal{O}.\texttt{send}(1^\lambda, P, \text{RAM}_0), \texttt{aux}_0\big) \stackrel{c}{\approx} \text{Sim}_{R\mathcal{O}}^{R\mathcal{O}_{P,\text{RAM}_0}}\big(1^\lambda, \texttt{aux}_0\big)$$
for auxiliary input $\texttt{aux}$ where we define $\text{Sim}_{R\mathcal{O}}^{R\mathcal{O}_{P,\text{RAM}_0}}$ as a collection of BQP algorithms $A_0, A_1, ..., A_\ell$ where

$\text{Sim}_{R\mathcal{O}}^{R\mathcal{O}_{P,\text{RAM}_0}} =$
$$A_\ell \circ (\mathbb{I}, R\mathcal{O}[P, \text{RAM}_{\ell-1}]) \circ ... \circ (\mathbb{I}, R\mathcal{O}[P, \text{RAM}_1]) \circ A_1 \circ (\mathbb{I}, R\mathcal{O}[P, \text{RAM}_0]) \circ A_0$$

and $(\mathbb{I}, R\mathcal{O}[P, \text{RAM}_{\ell-1}])$ represents a call to the RAM-oracle $R\mathcal{O}$ with initial state $\text{RAM}_{\ell-1}$ and identity oracle $\mathbb{I}$ (for passing state between $A_i, A_{i+1}$).

## 5.1 Instantiating RAM-Oracles with C-OTPs

We now show how to instantiate RAM-oracles using the semi-quantum one-time programs defined above (Definition 4.1).

First, we must define a "recursive" encoding for a program such that its evaluation produces a new one-time program to "refresh" the original program and allow for multiple queries to the program. We do this in a very similar manner to how Goyal et al. construct "pay-per-use" programs from one-time programs [21].

*Definition 5.3 (Recursive Program Encoding, $\overline{P}$)*: We define the program $\overline{P}_{\text{RAM}_i}^{\text{sk}}$ for original program $P : \mathcal{RAM} \times \mathcal{X} \rightarrow \mathcal{RAM} \times \mathcal{X}$ as follows:

$$\overline{P}^{\text{sk}}_{\text{RAM}_i}(x, \text{tag}_{i+1}) = (P(\text{RAM}_i, x), \text{ek}_{i+1}[P_{i+1}], \text{pk}_{i+2})$$

where:
- $\text{sk}_i = \text{PRF}(\text{sk}, i)$ and $\text{pk}_i = \text{C-OTP.setup}(\text{sk}_i)$
- $\text{ek}_{i+1}[P_{i+1}] = \mathcal{C}_{\text{C-OTP}}\left(\overline{P}_{\text{RAM}_{i+1}}, \text{sk}_{i+1}, \text{tag}\right)$ for the program $\overline{P}_{\text{RAM}_{i+1}}$ where $\text{RAM}_{i+1}, y = P(\text{RAM}_i, x)$

---

**Protocol 3:** RAM-Oracle Setup Scheme

---

**Sending Procedure**, $\text{R}\mathcal{O}.\text{send}(1^\lambda, P, \text{RAM}_0)$:

> The sender runs $\text{aux}_{\text{C-OTP}} \leftarrow \text{C-OTP.global\_setup}(\text{msk})$ for random secret key
>
> The sender chooses a program $P$ and initial RAM state $\text{RAM}_0$.
>
> The sender samples a secret key $\text{sk}$. Let $\text{sk}_0 = \text{PRF}(\text{sk}, 0)$ and $\text{sk}_1 = \text{PRF}(\text{sk}, 1)$.
>
> The sender generates
> $$\text{pk}_0 \leftarrow \text{C-OTP.setup}(\text{sk}_0, \text{aux}_{\text{C-OTP}}), \text{pk}_1 \leftarrow \text{C-OTP.setup}(\text{sk}_1, \text{aux}_{\text{C-OTP}}) \qquad (10)$$
> and oracle $\mathcal{O}_{\text{C-OTP}}$ with $\text{msk}$ embedded.
>
> The sender then sends $\text{pk}_0, \text{pk}_1, \mathcal{O}_{\text{C-OTP}}$ to the receiver.
>
> The sender and receiver engage in an interactive protocol specified by $\text{C-OTP.gen}$ for program $\overline{P}_{\text{RAM}_0}$ where $\overline{P}_{\text{RAM}_0}$ is defined as per Definition 5.3. The receiver ends up with evaluation key $\text{ek}_0$ for $\overline{P}_{\text{RAM}_0}$.

**Evaluation Procedure**, $\text{R}\mathcal{O}.\text{eval}\left(\overline{P}_{\text{RAM}_i}, x\right)$ for round $i \geq 0$:

> The receiver runs $|\text{👻}\rangle^{i+1}, \text{tag}_{i+1} \leftarrow \mathcal{Q}_{\text{C-OTP}}(\text{pk}_{i+1})$
>
> The receiver then runs $y, \text{ek}_{i+1}, \text{pk}_{i+2} \leftarrow \text{C-OTP.eval}\left((x, \text{tag}_{i+1}), |\text{👻}\rangle^i, \text{tag}_i, \text{ek}_i, \text{aux}_{\text{C-OTP}}\right)$ where $y = P(\text{RAM}_i, x)$
>
> The receiver then outputs result $y$ and stores the new state $|\text{👻}\rangle^{i+1}$, $\text{ek}_{i+1}$, and $\text{pk}_{i+2}$.

---

> *Theorem 5.1* (Scheme Simulation Soundness): The RAM-oracle scheme in Protocol 3 is sound assuming that the underlying C-OTP scheme is sound for program $\overline{P}_{\text{RAM}}$ for all $\text{RAM} \in \mathcal{RAM}$ as per Definition 4.2 as long as
> $$H_{\text{HILL}}(\text{sk} \mid \text{aux}_0, y_1, y_2, ..., y_\ell) = H_{\text{HILL}}(\text{sk})$$
> where $\text{aux}_0$ is the initial auxiliary information and $y_1, y_2, ..., y_\ell$ are the outputs of the program $P$ on queries $x_1, x_2, ..., x_\ell$ with RAM states $\text{RAM}_1, \text{RAM}_2, ..., \text{RAM}_\ell$[6].

*Proof*: Note that though the sending procedure is interactive, there is only one-round of interaction between the sender and receiver. So, we can model the adversary as a single algorithm, $\mathcal{A}$, with access to a *single classical* invocation of the sender's classical program which we will denote as $1\text{-}\mathcal{C}_{\text{C-OTP}}$. Also, as we give $\mathcal{A}$ access to oracle $\mathcal{O}_{\text{C-OTP}}$, we will write $\mathcal{A}^{1\text{-}\mathcal{C}_{\text{C-OTP}}, \mathcal{O}_{\text{C-OTP}}}$ to denote the adversary with access to the oracle $1\text{-}\mathcal{C}_{\text{C-OTP}}$ and $\mathcal{O}_{\text{C-OTP}}$. When clear from context, we will drop the superscript for $\mathcal{O}_{\text{C-OTP}}$ for brevity.

---

[6] We require this condition to ensure that the secret key $\text{sk}$ is sufficiently hidden from the adversary throughout the execution of the protocol as otherwise, the adversary could learn the obfuscated program while the simulator would be unable to from the secret key $\text{sk}$ alone.

We will prove that our scheme is sound by showing that
$$\mathcal{A}^{\mathcal{O}_{\text{C-OTP}},\ 1\text{-}\mathcal{C}_{\text{C-OTP}}(\overline{P},\text{sk}_0,\cdot)} \stackrel{c}{\approx}$$
$$S_\ell \circ (\mathbb{I},\ 1\text{-}P(\text{RAM}_{\ell-1},\text{tag}) \circ \ldots \circ (\mathbb{I},\ 1\text{-}P(\text{RAM}_1,\cdot)) \circ S_1 \circ (\mathbb{I},\ 1\text{-}P(\text{RAM}_0,\cdot)) \circ S_0. \tag{11}$$

for simulators $S_0, \ldots S_\ell$. Consider the following hybrids:

- $\text{Hyb}_0$: the real protocol
- $\text{Hyb}_1$: replace $\mathcal{A}^{1\text{-}\mathcal{C}_{\text{C-OTP}}}(\text{pk}_0, \text{aux})$ with the simulator for the one-time program: $S^{1\text{-}\overline{P}}$. Then break up $S$ into 2, $S_0, S'_1, \hat{S}_1$ such that
$$S = \hat{S}_1(\text{pk}_2, \cdot) \circ S'^{1\text{-}\mathcal{C}_{\text{C-OTP}}(\overline{P}_{\text{RAM}_1}, \text{sk}_1, \cdot)}_1 \circ S_0^{(1\text{-}P(\text{RAM}_0, \cdot))}. \tag{12}$$

- $\text{Hyb}_i$ for $i \in \{2, \ldots, \ell'\}$ for some $\ell' \geq \ell$ choosen later in the proof. Replace the last pair of simulators, $\hat{S}_{i-1} \circ S'_{i-1}$ with $\hat{S}_{i-1} \circ S^{1\text{-}\overline{P}_{\text{RAM}_{i-1}}}$. Then, take $S^{1\text{-}\overline{P}_{\text{RAM}_{i-1}}}$ and replace it with three simulators, $S_{i-1}, S'_i, \hat{S}_i$ such that
$$S^{1\text{-}\overline{P}_{\text{RAM}_{i-1}}} = \hat{S}_i(\text{pk}_{i+1}, \cdot) \circ S'^{1\text{-}\mathcal{C}_{\text{C-OTP}}(\overline{P}_{\text{RAM}_1}, \text{sk}_1, \cdot)}_i (\text{pk}_i, \cdot) \circ S_{i-1}^{(1\text{-}P(\text{RAM}_0, \cdot))}. \tag{13}$$

To prove that the hybrids are valid, we first note that as $H_{\text{HILL}}(\text{sk} \mid P, \text{aux}) = H_{\text{HILL}}(\text{sk})$, we have that $\text{sk}_i = \text{PRF}(\text{sk}, i)$ is indistinguishable from random and thus we can apply our simulation secure one-time programs.

Next, note that $\overline{P}_{\text{RAM}_i}$ simultaneously evaluates $\mathcal{C}_{\text{C-OTP}}(\overline{P}_{\text{RAM}_{i+1}}, \text{sk}_{i+1}, \cdot)$, outputs $\text{pk}_{i+2}$, and evaluates $P(\text{RAM}_i, \cdot)$. So, a simulator which can first evaluate $P(\text{RAM}_i, \cdot)$ and then evaluate $\mathcal{C}_{\text{C-OTP}}(\overline{P}_{\text{RAM}_{i+1}})$, and then receive $\text{pk}_{i+2}$ is strictly stronger than (and can thus simulate) the adversary in the real protocol.

Finally, note that after the $\ell$-th hybrid, we have
$$S = \hat{S}_\ell(\text{pk}_\ell, \cdot) \circ S'^{1\text{-}\mathcal{C}_{\text{C-OTP}}(\overline{P}_{\text{RAM}_1}, \text{sk}_1, \cdot)}_\ell \circ S^{1\text{-}P(\text{RAM}_{\ell-1}, \cdot)}_{\ell-1} \circ \ldots \circ S^{1\text{-}P(\text{RAM}_1, \cdot)}_1 \circ S^{1\text{-}P(\text{RAM}_0, \cdot)}_0 \tag{14}$$

Given that $S$ runs in polynomial time, we can always find some polynomially large $\ell' \geq \ell$ such that $\hat{S}_\ell, S'_\ell$ is the null simulator as $S$'s runtime is consumed by $S_{\ell-1}, \ldots, S_0$.

We thus have our desired result as we can re-write $S$ as a set of RAM oracle calls:
$$S_\ell \circ (\mathbb{I}, \text{R}\mathcal{O}[P, \text{RAM}_{\ell-1}]) \circ \ldots \circ (\mathbb{I}, \text{R}\mathcal{O}[P, \text{RAM}_1]) \circ S_1 \circ (\mathbb{I}, \text{R}\mathcal{O}[P, \text{RAM}_0]) \circ S_0. \tag{15}$$

∎

## 5.2 Short-Lived States are Good Enough

Another useful property of our RAM-oracle scheme is that we *do not* require long-term storage of quantum states but rather require short-term (though still with sufficient fidelity) storage of quantum states.

Specifically, we have the following lemma:

**Lemma 5.1** *(Short-Lived States are Good Enough)*: Assume that non-entangled qubits have independent noise and that the evaluation of the RAM-obfuscated recursively encoded program, $\mathcal{O}_{\overline{P}_{\text{RAM}}}$ takes $\omega$ time to evaluate. Then, if we are using the RAM-oracle scheme in Protocol 3 alongside the C-OTP scheme in Protocol 2, we have that
$$\Pr[\text{failed evaluation after } \ell \text{ queries}] \leq 1 - \ell \cdot p \tag{16}$$
where $p$ is the probability of a fault occurring during a single C-OTP evaluation taking place over time-scale $\omega$ and $\ell$ is the number of queries to the program.

*Proof*: The proof follows from a direct inspection of the RAM-oracle scheme in Protocol 3 and the union bound. Note that for each round of the RAM-oracle, we have to evaluate 1 one-time program. Then, the lack of shared quantum resources between rounds and independence assumption of non-entangled hardware means that the probability of a fault occuring in each round is independent of the probability of a fault occuring in all proceeding rounds. So, a simply application of the union bound gives us our desired result. ∎

Finally, we get that our RAM-oracle scheme is fault-tolerant with a logarithmic multiplicative overhead using Corollary 4.1.

**Corollary 5.1** *(Logarithmic Overhead implies fault-tolerance)*: $O\left(\log\left(\frac{\ell n}{\varepsilon_{\text{corr}}}\right) \cdot \frac{1}{\delta^2}\right)$ overhead is sufficient for a protocol with probability of correctness $1 - \varepsilon_{\text{corr}}$.

*Proof*: Recall, by Corollary 4.1, that the `CQ-TOK` scheme in outlined in Section 3.1 has a multiplicative overhead of $O\left(\log\left(\frac{\ell n}{\varepsilon_{\text{corr}}}\right) \cdot \frac{1}{\delta^2}\right)$ for achieving correctness probability $1 - \frac{\varepsilon_{\text{corr}}}{\ell n}$. So, we have that the overall failure probability for the RAM-oracle scheme in Protocol 3 is $\frac{\varepsilon_{\text{corr}}}{\ell n} \cdot \ell n = \varepsilon_{\text{corr}}$. ∎

# 6 Applications of RAM Oracles

In this section, we discuss some applications of our RAM oracles. Given that RAM oracles imply stateless black-box oracles, we will not delve into the applications of stateless black-box oracles as they are already well-known and numerous.

## 6.1 Long-Lived One-Time Programs without Long-Lived Quantum Memory

Long-lived quantum memory is a very expensive resource, requiring large error-correcting overhead. Especially when considering non-Markovian noise, it is not clear as to the cost of such memory (say holding a qubit for a year).

Instead, we can use our RAM oracles to construct a quantum cryptographic scheme which does not require long-lived quantum memory though "simulates" the use cases of such long-lived memory.

As one-time programs can be used to construct many quantum cryptographic schemes (such as certifiable deletion, quantum tokens, etc.), we will show how to construct "long-lived" one-time memories from our RAM oracles which imply both classical and one-time programs.

We will prove soundness relative to the following simulation-based notion for one-time memories:

*Definition 6.1 (One-Time Memory)*: A one-time memory is a protocol between a sender and receiver which can be represented as a tuple of algorithms (`prepState`, `readState`) where:
- `prepState` is a probabilistic algorithm which takes as input $s_0, s_1 \in \mathcal{S}$ and outputs a quantum state $\rho$ as well as classical auxiliary information aux.
- `readState` is a (potentially probabilistic) algorithm which takes as input $\rho$, aux and $\alpha \in \{0, 1\}$ and outputs a classical string $s_\alpha$ with probability $1 - \varepsilon$.

*Definition 6.2 (Correctness)*: A one-time memory (`prepState`, `readState`) is said correct with probability $\varepsilon$ if for all $s_0, s_1 \in \mathcal{S}$, we have that

$$\Pr[s_\alpha = s_{\alpha'}] \geq 1 - \varepsilon \qquad (17)$$

where $s_\alpha = \mathtt{readState}(\mathtt{prepState}(s_0, s_1), \alpha)$.

We adopt the definition of soundness for one-time memories in Ref. [10].

*Definition 6.3 (Soundness)*: A one-time memory $(\mathtt{prepState}, \mathtt{readState})$ is said to be sound relative to an adversary, $\mathcal{A}$, which interacts with the protocol, if there exists a simulator Sim for every inverse sub-exponential $\gamma(\cdot)$ for every $s_0, s_1 \in \mathcal{S}$ such that Sim makes at most one query to $g^{s_0,s_1}: \{0,1\} \to \{s_0, s_1\}$ (where $g(\alpha) = s_\alpha$) and

$$\mathcal{A}(\mathtt{prepState}(1^\lambda, s_0, s_1)) \stackrel{\gamma(\lambda)}{\approx} \mathrm{Sim}^{g^{s_0,s_1}}(1^\lambda) \qquad (18)$$

where $\stackrel{\gamma(\lambda)}{\approx}$ denotes statistical distance of at most $\gamma(\lambda)$.

---

**Oracle 4:** RAM Oracle for long-lived one-time memory for $\{s_0, s_1\}$

**Initial State** $\mathrm{RAM}_0 = \bot$ and current state $\mathrm{RAM}_i = a$ for $a \in \{\bot, \top\}$
**Inputs:** $b \in \{0,1\} \cup \{\bot\}$

1 **If** $a = \bot$ and $b \neq \bot$ **then**
2 $\quad \mathrm{RAM}_{i+1} = \top$
3 $\quad$ **Return** $s_b$
4 **Else**
5 $\quad \mathrm{RAM}_{i+1} = a$
6 $\quad$ **Return** $\bot$

---

We define the protocol for one-time memory as follows:

---

**Protocol 5:** Long-Lived One-Time Memory

$\mathtt{OTM.prepState}(1^\lambda, s_0, s_1)$:
$\quad$ Let $C$ be the RAM program outlined in Oracle 4.
$\quad$ With $\mathrm{RAM}_0 = \bot$, engage in the interactive protocol between the reciever and sender via $\mathrm{R}\mathcal{O}.\mathtt{send}(1^\lambda, C, \mathrm{RAM}_0)$
$\mathtt{OTM.readState}\left(1^\lambda, \overline{C}_{\mathrm{RAM}_i}, \alpha\right)$:
$\quad$ Run $\mathrm{R}\mathcal{O}.\mathtt{eval}\left(\overline{C}_{\mathrm{RAM}_i}, \alpha\right)$

---

*Theorem 6.1* (One-Time Program Correctness): The one-time memory protocol outlined above (Protocol 5) is correct with probability $1 - \varepsilon_{\mathrm{corr}}$ where $\varepsilon_{\mathrm{corr}}$ is the probability of error for the RAM oracle.

*Proof*: The proof follows from the fact that the RAM oracle is simulation sound and thus allows evaluation of $s_0$ and $s_1$ can be done with probability $1 - \varepsilon_{\mathrm{corr}}$. ∎

> *Theorem 6.2* (One-Time Program Security): For every BQP $\mathcal{A}$, the protocol outlined in Oracle 4 is sound.

*Proof*: Assume towards contradiction that the scheme is not simulation secure: i.e. there exists a BQP adversary $\mathcal{A}$ such that eq. (18) is broken. Then, we can construct a BQP adversary $\mathcal{A}'$ which breaks the simulation soundness of the RAM oracle. Specifically, $\mathcal{A}'$ calls $\mathcal{A}$ to get $s_0, s_1$. Then, $\mathcal{A}'$ can trivially distinguish between the real and RAM obfuscated simulated protocol as no set of calls to the RAM oracle returns a message other than $\bot$ more than once. ∎

## 6.2 Semi-Quantum Copy Protection

Though various definitions for copy-protection exist, we take the definition of copy-protection from Coladangelo et al. [26] due to both its simplicity and focus on classical circuits. We slightly modify the definition though to allow for the sender to be a classical party and the receiver to be a quantum party. We thus need to modify the setup (known as the "protection" procedure) to allow for an interactive protocol between the sender and receiver. We also allow for multi-bit output.

*Definition 6.4 (Quantum copy-protection scheme, [26])*: Let $\mathcal{C}$ be a family of classical circuits with an $m$ bit output. A quantum copy-protection (CP) scheme for $\mathcal{C}$ is a pair of QPT algorithms (CP.Protect, CP.Eval) with the following properties:

- CP.Protect $= \langle \text{Sen}(1^\lambda, C), \text{Rec} \rangle_{\langle \cdot, \text{OUT}_{\text{Rec}} \rangle}$: is a classical communication protocol between PPT Sen and QPT Rec.

At the end of the interaction Rec outputs a quantum state $\rho$.
- CP.Eval takes as input a quantum state $\rho$ and a string $x$, and outputs $m$ bits.

We say that the scheme is correct if, for any $\lambda \in \mathbb{N}$, $C \in \mathcal{C}$, and any input string $x$ to $C$:
$$\Pr[\text{CP.Eval}(\rho, x) = C(x) : \rho \leftarrow \text{CP.Protect}(1^\lambda, C)] \geq 1 - \text{negl}(\lambda). \qquad (19)$$

Ref. [26] then go on to define security as a game between a challenger and an adversary consisting of a triple of QPT algorithms $\mathcal{A} = (\mathcal{P}, \mathcal{F}_1, \mathcal{F}_2)$—a "pirate" $\mathcal{P}$ and two "freeloaders" $\mathcal{F}_1$ and $\mathcal{F}_2$. The game is specified by a security parameter $\lambda$, a distribution $D_\lambda$ over circuits in $\mathcal{C}$, and an ensemble $\{D_C\}_{\{C \in \mathcal{C}\}}$ where $D_C$ is a distribution over pairs of inputs to $C \in \mathcal{C}$. Moreover, they refer to $\{D_\lambda\}_{\lambda \in \mathbb{N}}$ as the program ensemble, and to $\{D_C\}_{C \in \mathcal{C}}$ as the input challenge ensemble. Then, the security game is defined as follows:

1. The challenger samples $C \leftarrow D_\lambda$ and sends $\rho \leftarrow \text{CP.Protect}(1^\lambda, C)$ to $\mathcal{P}$.
2. $\mathcal{P}$ creates a state on registers $A$ and $B$, and sends $A$ to $F_1$ and $B$ to $F_2$.
3. (input challenge phase:) The challenger samples $(x_1, x_2) \leftarrow D_C$ and sends $x_1$ to $F_1$ and $x_2$ to $F_2$. ($F_1$ and $F_2$ are not allowed to communicate).
4. $\mathcal{F}_1$ and $\mathcal{F}_2$ each return bits $b_1$ and $b_2$ to the challenger.

$A = (P, F_1, F_2)$ win if $b_1 = C(x_1)$ and $b_2 = C(x_2)$. Then, let random variable $\texttt{PiratingGame}(\lambda, P, F_1, F_2, D_\lambda, \{D_C\})$ denote whether the game is won.

Colagangelo et al. also define $p^{\text{triv}}_{D_\lambda,\{D_C\}_{C\in\mathcal{C}}}$ to be the winning probability that is trivially possible due to correctness: the pirate forwards the copy-protected program to one of the freeloaders, and leaves the other one with guessing as his best option. Formally:

*Definition 6.5 (The trivial probability of winning, $p^{\text{triv}}_{D_\lambda,\{D_C\}_{C\in\mathcal{C}}}$):* Let $\hat{D}_C$ be the induced distribution of winning answer pairs, and let $\hat{D}_{C,i}$, for $i \in \{1,2\}$ be its marginals. Then, one can define the optimal guessing probability of any of the two freeloaders,

$$p^{\text{triv}}_{D_\lambda,\{D_C\}_{C\in\mathcal{C}}} = \max_{i\in\{1,2\}} \max_{b\in\{0,1\}^m} \mathbb{E}_{\{C \leftarrow D_\lambda\}}\left[\hat{D}_{C,i}(b)\right]. \tag{20}$$

We can now define the security of a quantum copy-protection scheme.

*Definition 6.6 (Quantum copy-protection security, [26]):* A quantum copy-protection scheme for a family of circuits $\mathcal{C}$ is said to be secure with respect to the ensemble $\{D_\lambda\}_{\{\lambda\in\mathbb{N}\}}$ of distributions over circuits in $\mathcal{C}$, and with respect to the ensemble $\{D_C\}_{\{C\in\mathcal{C}\}}$, where $D_C$ is a distribution over pairs of inputs to program $C \in \mathcal{C}$, if for any QPT adversary $(P, F_1, F_2)$, any $\lambda \in \mathbb{N}$,

$$\Pr[\texttt{PiratingGame}(\lambda, P, F_1, F_2, D_\lambda, \{D_C\}) = 1] \leq p^{\text{triv}}_{D_\lambda,\{D_C\}_{C\in\mathcal{C}}} + \text{negl}(\lambda). \tag{21}$$

---

**Oracle 6:** RAM Oracle for Copy Protection for Classical Circuit, $C : \{0,1\}^n \to \{0,1\}^m$ and hardcoded PRF secret $K$

---

**Inital State** $\text{RAM}_0 = 0$
**Inputs** $\text{RAM}_i = i$ and input $x \in \{0,1\}^n, t \in \{0,1\}^\lambda$

1 **If** $i = -1$ **then**
2   | **Return** $\perp$
3 **If** $t \neq \texttt{PRF}(K, i)$ **then**
4   | **Set** $\text{RAM}_{i+1} = -1$
5   | **Return** $\perp$
6 **Else**
7   | **Set** $\text{RAM}_{i+1} = i + 1$
8   | **Return** $C(x), \texttt{PRF}(K, (i+1))$

---

**Protocol 7:** Copy Protection from RAM Oracles

---

$\texttt{CP.Protect}(1^\lambda, C)$:
  | Let $C'$ be the RAM program outlined in Oracle 6.
  | With $\text{RAM}_0 = 0$, engage in the interactive protocol between the reciever and sender via
  | $\text{R}\mathcal{O}.\texttt{send}(1^\lambda, C', \text{RAM}_0)$
$\texttt{CP.Eval}(1^\lambda, \overline{C}_{\text{RAM}_i}, x)$:
  | Run $\text{R}\mathcal{O}.\texttt{eval}(\overline{C}_{\text{RAM}_i}, x)$

---

We can then use our RAM oracle to construct a copy-protection scheme as shown in Protocol 7.

*Theorem 6.3* (Copy Protection Correctness): The protocol in Protocol 7 is correct with probability $\varepsilon_{\text{corr}}$ where $\varepsilon_{\text{corr}}$ is the probability of correctness for the RAM oracle.

*Proof*: The proof follows from the fact that RAM obfuscation allows for the simulation of the program $C$ and thus evaluation of $C(x)$. ∎

*Theorem 6.4* (Copy Protection Security): For every BQP $\mathcal{A}$, the above protocol is sound: i.e.
$$\Pr[\texttt{PiratingGame}(\lambda, \mathcal{P}, \mathcal{F}_1, \mathcal{F}_2, D_\lambda, \{D_C\}) = 1] \le p^{\text{triv}}_{D_\lambda, \{D_C\}_{C \in \mathcal{C}}} + \text{negl}(\lambda). \tag{22}$$

*Proof*: Assume towards contradiction that the scheme is not copy-protection secure. Then, there exists freeloaders, $\mathcal{F}_1, \mathcal{F}_2$ such that $\mathcal{F}_1$ and $\mathcal{F}_2$ can output $P(x_1), P(x_2)$ respectively without communicating. Then, we can construct a BQP adversary $\mathcal{A}'$ which breaks the simulation soundness of the RAM oracle. First, note that by the soundness of RAM obfuscation, we can consider the collection $(\mathcal{P}, \mathcal{F}_1, \mathcal{F}_2)$ as a single adversary, $\mathcal{A}'$ which is then modeled by a tuple of algorithms, $A_1, ..., A_q$, for some polynomial $q$, such that
$$\mathcal{A}' = A_q \circ \text{R}\mathcal{O}[C, \text{RAM}_{q-1}] \circ ... \circ \text{R}\mathcal{O}[C, \text{RAM}_1] \circ A_1 \circ \text{R}\mathcal{O}[C, \text{RAM}_0] \circ A_0. \tag{23}$$

Next, note that if $\mathcal{F}_1$ and $\mathcal{F}_2$ can output $P(x_1), P(x_2)$ beyond the trivial probability, then $\mathcal{A}'$ must make a call to the RAM oracle with $x_1$ and $x_2$. We will say that $x$ is called at time step $i$ if $\mathcal{A}'$ makes a call to the RAM oracle with $x_\alpha$ at time step $i$ (i.e. with $A_i$). Note that by simulation soundness, only one call to the RAM oracle can be made at a time step. Then, there are two cases:

**Case 1**: $x_1$ and $x_2$ are called at the same time step. Note that as $\mathcal{F}_1$ and $\mathcal{F}_2$ do not communicate, then $\tilde{P}$ must be called *separately* at the same time step. But note that, by the simulation soundness of RAM obfuscation, $\mathcal{A}'$ can only make *one* call to the RAM oracle at a time step.

**Case 2**: $x_1$ and $x_2$ are called at different time steps. Assume that $x_1$ is called at timestep $i$ and $x_2$ is called at timestep $j$ with $i < j$. But then, as $\mathcal{F}_1$ and $\mathcal{F}_2$ do not communicate, $\mathcal{F}_1$ has some state after timestep $i$ which contains $\texttt{PRF}(K, i)$. Note that as $\mathcal{F}_2$ cannot call the RAM oracle at timestep $i$ and that $\texttt{PRF}(K, i) \stackrel{c}{\approx}$ uniform, $\mathcal{F}_2$ cannot call the RAM oracle at timestep $i + 1$ without setting the state to the constant reject state (i.e. the state set to $-1$). But then, $\mathcal{F}_2$ cannot call the RAM oracle at timestep $i + 2, ..., i + (j - i)$ as only $\mathcal{F}_1$ can call the RAM oracle at these time steps.

And so, if $\mathcal{F}_1$ and $\mathcal{F}_2$ can output $P(x_1), P(x_2)$ with non-trivial probability, then $\mathcal{A}'$ must break the simulation soundness of the RAM oracle or $\mathcal{F}_1$ and $\mathcal{F}_2$ must communicate. ∎

**Remark 6.1**: Though we do not explore *stateful copy-protection* in this paper, we note RAM obfuscation seems amenable to this task as well.

### 6.3 More Applications

RAM obfuscation is a powerful primitive which can be used to construct a variety of other primitives. Without going into details, we can also construct:

- Smart contracts and cryptocurrency without a blockchain: just as in Amos et al. [27], we can construct a blockchain-less cryptocurrency by having a RAM oracle act as a "trusted wallet" which keeps track of a user's balance. When the user want to send some money, they use their RAM oracle (as a trusted wallet) to interact with the RAM oracle of the receiver. Then, both RAM oracles update their state to reflect the new balance. We can view this as a sort of "strengthening" of quantum money.
- Obfuscation for turing machines and processors: the program of the RAM oracle can be used to update the state of a turing machine or procesor at each step in a rather simple way.
- A "unique soul" for a computer: though a sci-fi-esque concept, we can use copy protection alongside RAM oracles to ensure that a robot cannot copy itself but also cannot be internally inspected! In some sense, this can be thought of as a "unique soul" for a computer in that it is bound to one point in space and time but cannot be inspected and modified at will.

## 7 Conclusions and Future Work

In this work, we have shown how to construct a variety of quantum cryptographic primitives using classically accessible oracles and semi-quantum tokens. Using the semi-quantum nature of quantum tokens, we are able to construct semi-quantum one-time programs and RAM obfuscation schemes which inherit the underlying fault-tolerance of the quantum tokens. We also show how to use our RAM obfuscation scheme to construct long-lived one-time programs and copy-protection schemes: i.e. we construct primitives which can be used over long time periods *without* requiring coherent quantum memory or global entanglement over the same time period.

At first glance, our use of publicly-verifiable quantum tokens may be too strong of a requirement for our construction as we only allow for the adversary to have *classical* access to token's verification oracle. We thus believe that our construction can be simplified to use weaker cryptographic primitives, such as remote-state preperation in combination with quantum MAC tokens [28], which have the added benefit of being more efficient and noise tolerant. We leave this as an open problem for future work.

We also note that though our construction is "fault-tolerant," if the underlying quantum token scheme has correctness error less than half, achieving a non-trivial correctness error is quite a difficult task. Though the progress of quantum error correction and computation has been rapid, the construction of non-trivial correctness error for quantum tokens remains a challenging open problem.

## Acknowledgements

The author is grateful to the helpful discussions and feedback from Fabrizio Romano Genovese, Gorjan Alagic, Stefano Gogioso, and Yi-Kai Liu. The author also acknowledges funding and support from NeverLocal Ltd, Neon Tetra LLC, and from the NSF Graduate Research Fellowship Program.

# Bibliography

[1] S. Goldwasser, Y. T. Kalai, and G. N. Rothblum, *One-Time Programs*, in *Advances in Cryptology–CRYPTO 2008: 28th Annual International Cryptology Conference, Santa Barbara, CA, USA, August 17-21, 2008. Proceedings 28* (2008), pp. 39–56.

[2] S. Gunn and R. Movassagh, Quantum one-time protection of any randomized algorithm, Arxiv Preprint Arxiv:2411.03305 (2024).

[3] A. Gupte, J. Liu, J. Raizes, B. Roberts, and V. Vaikuntanathan, Quantum one-time programs, revisited, Arxiv Preprint Arxiv:2411.01876 (2024).

[4] M.-C. Roehsner, J. A. Kettlewell, J. Fitzsimons, and P. Walther, Probabilistic one-time programs using quantum entanglement, Npj Quantum Information **7**, 98 (2021).

[5] A. Broadbent, S. Gharibian, and H.-S. Zhou, Towards Quantum One-Time Memories from Stateless Hardware, Quantum **5**, 429 (2021).

[6] A. Behera, O. Sattath, and U. Shinar, Noise-tolerant quantum tokens for MAC, Arxiv Preprint Arxiv:2105.05016 (2021).

[7] K.-M. Chung, M. Georgiou, C.-Y. Lai, and V. Zikas, Cryptography with disposable backdoors, Cryptography **3**, 22 (2019).

[8] L. Stambler, Quantum One-Time Memories from Stateless Hardware, Random Access Codes, and Simple Nonconvex Optimization, Arxiv Preprint Arxiv:2501.04168 (2025).

[9] L. Stambler, Information Theoretic One-Time Programs from Geometrically Local QNC0 Adversaries, Arxiv Preprint Arxiv:2503.22016 (2025).

[10] Q. Liu, *Depth-Bounded Quantum Cryptography with Applications to One-Time Memory and More.*, in *14th Innovations in Theoretical Computer Science Conference (ITCS 2023)* (2023).

[11] Y.-K. Liu, *Building One-Time Memories from Isolated Qubits*, in *Proceedings of the 5th Conference on Innovations in Theoretical Computer Science* (2014), pp. 269–286.

[12] Y.-K. Liu, *Single-Shot Security for One-Time Memories in the Isolated Qubits Model*, in *Advances in Cryptology–CRYPTO 2014: 34th Annual Cryptology Conference, Santa Barbara, CA, USA, August 17-21, 2014, Proceedings, Part II 34* (2014), pp. 19–36.

[13] A. J. Leggett, S. Chakravarty, A. T. Dorsey, M. P. Fisher, A. Garg, and W. Zwerger, Dynamics of the dissipative two-state system, Reviews of Modern Physics **59**, 1 (1987).

[14] M. H. Amin, P. J. Love, and C. Truncik, Thermally assisted adiabatic quantum computation, Physical Review Letters **100**, 60503 (2008).

[15] M. Amin and F. Brito, Non-Markovian incoherent quantum dynamics of a two-state system, Physical Review B—Condensed Matter and Materials Physics **80**, 214302 (2009).

[16] M. W. Johnson et al., Quantum annealing with manufactured spins, Nature **473**, 194 (2011).

[17] C. Gidney and M. Ekerå, How to factor 2048 bit RSA integers in 8 hours using 20 million noisy qubits, Quantum **5**, 433 (2021).

[18] S. Aaronson, *Quantum Copy-Protection and Quantum Money*, in *2009 24th Annual IEEE Conference on Computational Complexity* (2009), pp. 229–242.

[19] S. Aaronson, J. Liu, Q. Liu, M. Zhandry, and R. Zhang, *New Approaches for Quantum Copy-Protection*, in *Advances in Cryptology–CRYPTO 2021: 41st Annual International Cryptology Conference, CRYPTO 2021, Virtual Event, August 16–20, 2021, Proceedings, Part I 41* (2021), pp. 526–555.

[20] O. Shmueli, *Semi-Quantum Tokenized Signatures*, in *Annual International Cryptology Conference* (2022), pp. 296–319.

[21] R. Goyal and V. Goyal, *Overcoming Cryptographic Impossibility Results Using Blockchains*, in *Theory of Cryptography: 15th International Conference, TCC 2017, Baltimore, MD, USA, November 12-15, 2017, Proceedings, Part I 15* (2017), pp. 529–561.

[22] M. Bellare and A. Sahai, *Non-Malleable Encryption: Equivalence between Two Notions, And an Indistinguishability-Based Characterization*, in *Annual International Cryptology Conference* (1999), pp. 519–536.

[23] D. Boneh and V. Shoup, A graduate course in applied cryptography, Draft 0.5 (2020).

[24] J. Håstad, R. Impagliazzo, L. A. Levin, and M. Luby, A pseudorandom generator from any one-way function, SIAM Journal on Computing **28**, 1364 (1999).

[25] C.-Y. Hsiao, C.-J. Lu, and L. Reyzin, *Conditional Computational Entropy, Or toward Separating Pseudoentropy from Compressibility*, in *Advances in Cryptology-EUROCRYPT 2007: 26th Annual International Conference on the Theory and Applications of Cryptographic Techniques, Barcelona, Spain, May 20-24, 2007. Proceedings 26* (2007), pp. 169–186.

[26] A. Coladangelo, C. Majenz, and A. Poremba, Quantum copy-protection of compute-and-compare programs in the quantum random oracle model, Quantum **8**, 1330 (2024).

[27] R. Amos, M. Georgiou, A. Kiayias, and M. Zhandry, One-shot Signatures and Applications to Hybrid Quantum/Classical Authentication, (2020).

[28] A. Behera, O. Sattath, and U. Shinar, Noise-Tolerant Quantum Tokens for MAC, (2021).

# A Modified Token Generation Protocol

We present a modified token generation protocol for the semi-quantum tokenized signature scheme. As the signing and verification algorithm remain completely unchanged from [20], we only present the modified token generation protocol.

---

**Protocol 8:** Token Generation Protocol

---

Sen is classical and Rec is quantum. The joint input is the security parameter $\lambda \in \mathbb{N}$.

Setup:

1. Sen samples a random $\frac{\lambda}{2}$-dimensional subspace $S \subset \{0,1\}^\lambda$, described by a matrix $M_S \in \{0,1\}^{\frac{\lambda}{2} \times \lambda}$
2. Sen samples OTP key $p_x \leftarrow \{0,1\}^{\frac{\lambda^2}{2}}$ to encrypt $M_S^{(p_x)} = \text{QHE.OTP}_{p_x}(M_S)$
3. Sen generates fhek $\leftarrow \text{QHE.Gen}(1^\lambda, 1^{\ell(\lambda)})$ for some polynomial $\ell(\cdot)$
4. Sen computes $\text{ct}_{p_x} \leftarrow \text{QHE.Enc}_{\text{fhek}}(p_x)$
5. Sen sends the encryption $\text{pk} = \left(M_S^{(p_x)}, \text{ct}_{p_x}\right)$ to Rec

$\text{Rec}\left(M_S^{(p_x)}, \text{ct}_{p_x}\right)$:

6. Let $C$ be the quantum circuit that for an input matrix $M \in \{0,1\}^{\frac{\lambda}{2} \times \lambda}$, outputs a uniform superposition of its row span
7. Rec homomorphically evaluates $C$: $\left(|S\rangle_{x,z}, \text{ct}_{x,z}\right) \leftarrow \text{QHE.Eval}\left(\left(M_S^{(p_x)}, \text{ct}_{p_x}\right), C\right)$
8. Rec saves the quantum part $|S\rangle_{x,z}$ and sends the classical part $\text{tag} = \text{ct}_{x,z}$ to Sen

Sen then:

9. Sen decrypts tag using the OTP key $p_x$ to get $(x, z) = \text{QHE.Dec}_{p_x}(\text{tag})$
10. **if** $x \in S$ **then**
11. ┃ The interaction is terminated
12. Let $M_{S^\perp} \in \{0,1\}^{\{\frac{\lambda}{2} \times \lambda\}}$ be a basis for $S^\perp$ (as a matrix)
13. Let $w$ be the first row in $M_S$
14. Let $M_{S_0} \in \{0,1\}^{\{(\frac{\lambda}{2}-1) \times \lambda\}}$ be the rest of the matrix $M_S$, without $w$
15. Sen computes indistinguishability obfuscations:
16. ┃ $O_{S_0+x} \leftarrow \text{iO}(M_{S_0}, x)$
17. ┃ $O_{S_0+w+x} \leftarrow \text{iO}(M_{S_0}, w+x)$
18. ┃ $O_{S^\perp+z} \leftarrow \text{iO}(M_{S^\perp}, z)$
19. All obfuscations use padding poly'($\lambda$) for some polynomial poly'
20. Sen sends the obfuscations to Rec

---

# B Missing Proofs for the Semi-Quantum OTP

In this section, we present the missing proofs for the semi-quantum OTP scheme. We first recall the series of hybrid games used in the proof of Theorem 4.1:

- $\text{Hyb}_0$: the real protocol
- $\text{Hyb}_{1,0}$: replace the first call to $\mathcal{O}_{\text{C-OTP}}$ with $\mathcal{O}_1(x, \text{ct}', \sigma)$ as follows:

- If $P'$ is independent of $P$ (i.e. $H_{\mathsf{HILL}}(P \mid P') = H_{\mathsf{HILL}}(P)$ and vice-versa), return $\mathcal{O}_{\mathsf{C\text{-}OTP}}(x, \mathsf{ct}', \sigma)$
- If $\mathsf{ct}' = \mathsf{ct}_A$, return $\mathcal{O}_{\mathsf{C\text{-}OTP}}(x, \mathsf{ct}_A, \sigma)$
- Otherwise, return $\bot$

- $\mathsf{Hyb}_{1,i}$: replace the $i$-th call to $\mathcal{O}_{\mathsf{C\text{-}OTP}}$ with $\mathcal{O}_1\left(x, \mathsf{ct}'_{P', \mathsf{pk}', \mathsf{ek}'}, \sigma\right)$ as follows:
  - If $P'$ is independent of $P$ given the prior $i-1$ calls to $\mathcal{O}_1$, return $\mathcal{O}_1(x, \mathsf{ct}', \sigma)$
  - If $\mathsf{ct}' = \mathsf{ct}_A$, return $\mathcal{O}_{\mathsf{C\text{-}OTP}}(x, \sigma, \mathsf{ct}_A, \mathsf{ek}')$
  - Otherwise, return $\bot$

- $\mathsf{Hyb}_2$: Replace $\mathsf{ct}_A = \mathsf{ct}_{P, \mathsf{sk}}$ with $\mathsf{ct}_\bot = \mathsf{PK.Encr}(\mathsf{mpk}, [\bot, \mathsf{pk}, \mathsf{ek}])$. Then, replace calls to $\mathcal{O}_1(x, \mathsf{ct}', \sigma)$ with $\mathcal{O}_2(x, \mathsf{ct}', \sigma)$ as follows:
  - If $\mathsf{ct}' = \mathsf{ct}_\bot$, return $\mathcal{O}_1(x, \mathsf{ct}_A, \sigma)$
  - Otherwise, return $\mathcal{O}_1(x, \mathsf{ct}', \sigma)$

- $\mathsf{Hyb}_3$: Replace $\mathcal{O}_2$ with $\mathcal{O}_3^{1\text{-}P}(x, \mathsf{ct}', \sigma)$ as follows:
  - If $\mathsf{ct}' \neq \mathsf{ct}_A$, return $\mathcal{O}_2(x, \mathsf{ct}', \sigma)$
  - Check if $\mathsf{CQ\text{-}TOK.CV}(\mathsf{pk}^i_{\mathsf{CQ\text{-}TOK}}, \mathsf{ek}_i, \sigma_i, x_i) = 1$ for all $i \in [n]$.
    – If the check passed, then check if $x$ has been called and if yes, retrieve $P(x)$ from memory. If not, return $P(x)$ and store $P(x)$. Note that each check passes with probability at least $1 - \varepsilon_{\mathsf{corr}}$ and thus, for honest generation, the check passes with probability at least $1 - n \cdot \varepsilon_{\mathsf{corr}}$.
    – Otherwise, return $\bot$

- $\mathsf{Hyb}_4$: The same as before except that the simulator samples random public key and private key pairs for the quantum tokens, $\widehat{\mathsf{pk}}_i, \widehat{\mathsf{sk}}_i$. Then sample a corresponding evaluation key, $\widehat{\mathsf{ek}}_i$, for the tag. Replace the encryption of $\mathsf{ct}_\bot$ with $\widehat{\mathsf{ct}}_\bot = \mathsf{PK.Encr}\left(\mathsf{mpk}, \left[\bot, \widehat{\mathsf{pk}}, \widehat{\mathsf{ek}}\right]\right)$

- $\mathsf{Hyb}_5$: Replace the adversary with the simulator

**Lemma 6.1**:
$$\mathsf{Hyb}_0 \stackrel{c}{\approx} \mathsf{Hyb}_{1,0} \tag{24}$$

and

$$\mathsf{Hyb}_{1,i-1} \stackrel{c}{\approx} \mathsf{Hyb}_{1,i} \tag{25}$$

as long as $\mathcal{A}$ makes at most polynomially many queries to $\mathcal{O}_{\mathsf{C\text{-}OTP}}$.

*Proof*: Assume towards contradiction that there exists a distinguisher $D$ which can distinguish between $\mathsf{Hyb}_0$ and $\mathsf{Hyb}_{1,0}$ or $\mathsf{Hyb}_{1,i}$ and $\mathsf{Hyb}_{1,i+1}$ with non-negligible probability. The proof follows in the same manner for both cases, so we will assume that $D$ can distinguish between $\mathsf{Hyb}_{1,i}$ and $\mathsf{Hyb}_{1,i+1}$ with non-negligible probability. Then , let $s_2$ be the protocol's state after the $i$ queries to $\mathcal{O}_1$. Note that $s_2$ is independent of $\mathsf{msk}$ as we require that $H_{\mathsf{HILL}}(\mathsf{msk} \mid P, \mathsf{aux}) = H_{\mathsf{HILL}}(\mathsf{msk})$ and so neither any call to the program or auxilary state reveal information about $\mathsf{msk}$.

Next, recall the two indistinguishable games in the security definition of non-malleable encryption (Definition 2.1):

$$
\begin{array}{ll}
\texttt{Expt}_{\mathcal{A},\,\mathsf{PK},R}(\lambda)\text{:} & \texttt{Expt}_{\mathsf{Sim},\,\mathsf{PK},R}(k) \\
\quad (\mathtt{pk},\mathtt{sk}) \leftarrow \mathsf{PK.Gen}(1^\lambda) & \quad (\mathtt{pk},\mathtt{sk}) \leftarrow \mathsf{PK.Gen}(1^k) \\
\quad (M, s_1, s_2) \xleftarrow{\$} \mathcal{A}_1(\mathtt{pk}) & \quad (M, s_1, s_2) \leftarrow \mathrm{NMSim}_1(\mathtt{pk}) \\
\quad x \leftarrow M;\, y \xleftarrow{\$} \mathsf{PK.Encr}_{\mathtt{pk}}(x) & \quad x \leftarrow M \\
\quad y' \leftarrow \mathcal{A}_2(y, s_2) & \quad y' \leftarrow \mathrm{NMSim}_2(s_2) \\
\quad x' \leftarrow \mathsf{PK.Decr}_{\mathtt{sk}}(y') & \quad x' \leftarrow \mathsf{PK.Decr}_{\mathtt{sk}}(y') \\
\quad \text{If } R(x, x', M, s_1) \text{ then return } 1 & \quad \text{If } R(x, x', M, s_1) \text{ then return } 1 \\
\quad \text{Else return } 0 & \quad \text{Else return } 0 \quad\quad (26)
\end{array}
$$

where $\mathtt{sk}$ and $\mathtt{pk}$ correspond to the master secret key ($\mathtt{msk}$) and its associated public key ($\mathtt{mpk}$).

Then, in the non-malleable experiment (Definition 2.1), define relationship $R$ as follows:
$$
\begin{aligned}
& R(x = (P, \mathtt{ek}, \mathtt{pk}), x' = (P', \mathtt{ek}', \mathtt{pk}'), M, \text{""}) = 1 \\
& \quad \text{if } \Pr\big[P' = \tilde{P} \,\big|\, \tilde{P} \leftarrow \mathcal{B}(P, s_2)\big] - \Pr\big[P' = \tilde{P} \,\big|\, \tilde{P} \leftarrow \mathcal{B}'(s_2)\big] > \varepsilon
\end{aligned}
\quad (27)
$$

for all BQP algorithm $\mathcal{B}, \mathcal{B}'$, $\varepsilon \in \mathrm{negl}(\lambda)$. We now show that if $D$ can distinguish between $\mathtt{Hyb}_{1,i}$ and $\mathtt{Hyb}_{1,i+1}$, then we can construct an adversary $\mathcal{A}$ which can distinguish the real experiment from the simulated experiment of Definition 2.1 with non-negligible probability.

Note that $\mathrm{NMSim}_2$ in the non-malleable simulator which produces cipher-text encrypting $y' = (P', \mathtt{ek}', \mathtt{pk}')$ independently of cipher-text $y = \mathsf{PK.Encr}_{\mathtt{pk}}(P, \mathtt{ek}, \mathtt{pk})$ given $s_2$. Thus, $P'$ is independent of $P$. But, if $D$ can distinguish between $\mathtt{Hyb}_{1,i}$ and $\mathtt{Hyb}_{1,i+1}$, then $\mathcal{A}$ can distinguish between the real and simulated experiments with non-negligible probability as, in the real distribution, $D$ can find some ciphertext encoding $P'$ which is not independent of $P$. ∎

**Lemma 6.2**:
$$
\mathtt{Hyb}_{1,q} \stackrel{\mathrm{c}}{\approx} \mathtt{Hyb}_2 \quad (28)
$$

*Proof*: Assume towards contradiction that distinguisher $D$ can distinguish between the two hybrids. We can define $\mathcal{A}$ whichs breaks the CPA security of the public key encryption scheme as follows:

- First $\mathcal{A}$ receives the program $P$ as auxiliary input as well as the secret key $\mathtt{sk}$.
- Then, $\mathcal{A}$ simulates the distribution of $D$ in the real protocol given one of two cipher-texts $\mathtt{ct}_A$ or $\mathtt{ct}_\perp$.
- Then, $\mathcal{A}$ can distinguish between the two distributions with non-negligible probability using $D$.

Next, note that the following two distributions are indistinguishable by CPA
$$
\begin{array}{ll}
\texttt{Expt}_{\mathcal{A},\,\mathsf{PK},R}(\lambda)\text{:} & \texttt{Expt}_{\mathcal{A},\,\mathsf{PK},R}(\lambda)\text{:} \\
\quad (\mathtt{pk},\mathtt{sk}) \leftarrow \mathsf{PK.Gen}(1^\lambda) & \quad (\mathtt{pk},\mathtt{sk}) \leftarrow \mathsf{PK.Gen}(1^\lambda) \\
\quad A = (\mathtt{mpk}, [P, \mathtt{pk}, \mathtt{ek}]) \leftarrow \mathcal{A}_1(\mathtt{pk}) & \quad A = [P, \mathtt{pk}, \mathtt{ek}] \leftarrow \mathcal{A}_1(\mathtt{pk}) \\
\quad y \xleftarrow{\$} \mathsf{PK.Encr}_{\mathtt{pk}}(A) & \quad y \xleftarrow{\$} \mathsf{PK.Encr}_{\mathtt{pk}}([\perp, \mathtt{pk}, \mathtt{ek}]) \\
\quad y' \leftarrow \mathcal{A}_2(y, A) & \quad y' \leftarrow \mathcal{A}_2(y, A) \\
\quad \text{Return } y' & \quad \text{Return } y' \quad\quad (29)
\end{array}
$$

Given knowledge of $A = [P, \mathtt{ek}, \mathtt{pk}]$, $\mathcal{A}_2$ can simulate all calls to $\mathcal{O}_1$ and $\mathcal{O}_2$ in both cases. So, $\mathcal{A}_2$ can internally simulate distinguisher $D$. Then, if the $\mathcal{A}_2$ can distinguish between the two distributions, it can distinguish between the game in eq. 29 with non-negligible probability, breaking the cpa security of the public key encryption scheme. ∎

**Lemma 6.3**:
$$\text{Hyb}_2 \stackrel{c}{\approx} \text{Hyb}_3 \qquad (30)$$

*Proof*: Assume towards contradiction that a distinguisher $D$ can distinguish the two hybrids. Then, with non-negligible probability, $D$ can query $\mathcal{O}_2$ with $x, \text{ct}', \sigma$ and $x', \text{ct}', \sigma'$ with $x \neq x'$ such that CQ-TOK.CV outputs 1 for both queries. Then, $D$ can break the security of the token scheme as the adversary can produce a valid signature for $x$ and $x'$ under the same evaluation key. Note that CQ-TOK.ek is independent of $P, \text{ek}$. Without loss of generality, assume that $x'_i \neq x_i$.

Define $\mathcal{A}$ as follows:
- Sample random sk'
- Let $\text{pk} = (\text{pk}_1, ..., \text{pk}_n, \text{PK.Encr}(\text{mpk}, [\bot, \text{ek}', \text{pk}']))$ where $\text{pk}'_1, ..., \text{pk}'_{i-1}, \text{pk}_i, ..., \text{pk}'_n$ are the public keys for the token scheme with $\text{pk}'_j$ for $i \neq j$ equaling a ample of $\text{PK.Gen}(1^\lambda)$
- Internally simulate C-OTP.gen with pk except that we use the real messages for the $i$-th evaluation keys to get $(\text{ek}'_1, ..., \text{ek}'_{i-1}, \text{ek}_i, \text{ek}'_{i+1}, ..., \text{ek}'_n)$ with $i$ being randomly choosen.
- Output $x, \sigma, \text{ct}', \text{CQ-TOK.ek}$ and $x', \sigma', \text{ct}', \text{CQ-TOK.ek}$ with $x \neq x'$ using $D$.

Note that, in the above, $\mathcal{A}$ simulates the distribution of $D$ in the real protocol. Also, note that as $x \neq x'$, $\Pr_i[x_i \neq x'_i] \geq \frac{1}{n}$. And so, if $D$ can output two signatures for different messages such that CQ-TOK.CV outputs 1 with non-negligible probability, than $\mathcal{A}$ can break the security definition of Definition 3.1 with inverse polynomial probability. ∎

**Lemma 6.4**:
$$\text{Hyb}_3 = \text{Hyb}_4 \qquad (31)$$

*Proof*: Note that the cipher-text, $\text{ct}_\bot$, in the previous hybrid encodes a random public-key and the null program, $\bot$. Thus, $\widehat{\text{ct}}_\bot$ encodes the same distribution of underlying messages as $\text{ct}_\bot$. ∎

**Lemma 6.5**:
$$\text{Hyb}_4 = \text{Hyb}_5 \qquad (32)$$

*Proof*: Note that as the simulator samples $\widehat{\text{pk}}, \widehat{\text{sk}}, \widehat{\text{ek}}$ in lieu of the setup party, the simulator can now simulate the call to $\mathcal{C}_{\text{C-OTP}}$. Moreover, by $\text{Hyb}_3$, the simulator can make a single call to $P$ va $\mathcal{O}_3^{1\text{-}P}$. We thus have that the original adversary's view is indistinguishable to the simulator's, $\text{Sim}^{1\text{-}P, \mathcal{O}_{\text{C-OTP}}}(\text{aux})$. ∎